\documentclass[twocolumn]{aastex701}
\usepackage{xspace}
\usepackage{ulem}

\newcommand{\lya}{Ly$\alpha$\xspace}

\newcommand{\hei}{He~{\sc i}$\lambda$10830\xspace}
\newcommand{\pag}{Pa\,$\rm \gamma$\xspace}
\newcommand{\cf}{\texttt{cid\_414}}
\newcommand{\cn}{\texttt{cid\_947}}

\newcommand{\edd}{$\lambda_{\rm Edd}$}

\begin{document}


\title{COSMOS-3D: Dense Circumnuclear Gas across Black Hole Growth Phases at $z\sim3$}

\correspondingauthor{Siwei Zou}
\email{zousw@bao.ac.cn}

\author[0000-0001-7634-1547]{Zi-Jian Li}
\affiliation{Chinese Academy of Sciences South America Center for Astronomy, National Astronomical Observatories, CAS, Beijing 100101, China}
\affiliation{School of Astronomy and Space Sciences, University of Chinese Academy of Sciences, Beijing 100049, China}
\email{zjli@nao.cas.cn}

\author[0000-0002-3983-6484]{Siwei Zou}
\affiliation{Chinese Academy of Sciences South America Center for Astronomy, National Astronomical Observatories, CAS, Beijing 100101, China}
\affiliation{Departamento de Astronom\'ia, Universidad de Chile, Casilla 36-D, Santiago, Chile}
\email{zousw@nao.cas.cn}

\author[0000-0002-6221-1829]{Jianwei Lyu}
\affiliation{Steward Observatory, University of Arizona,933 N Cherry Ave, Tucson, AZ, 85721, USA}
\email{jianwei@arizona.edu }

\author[0000-0002-6184-9097]{Jaclyn B. Champagne}
\affiliation{Steward Observatory, University of Arizona,933 N Cherry Ave, Tucson, AZ, 85721, USA}
\email{jackie.champagne.astro@gmail.com}

\author[0000-0001-6511-8745]{Jia-Sheng Huang}
\affiliation{Chinese Academy of Sciences South America Center for Astronomy, National Astronomical Observatories, CAS, Beijing 100101, China}
\affiliation{Harvard-Smithsonian Center for Astrophysics, 60 Garden Street, Cambridge, MA 02138, USA}
\email{jhuang@nao.cas.cn }

\author[0000-0003-0202-0534]{Cheng Cheng}
\affiliation{Chinese Academy of Sciences South America Center for Astronomy, National Astronomical Observatories, CAS, Beijing 100101, China}
\email{chengcheng@nao.cas.cn}

\author[0000-0003-0964-7188]{Shuqi Fu}
\affiliation{Kavli Institute for Astronomy and Astrophysics, Peking University, Beijing 100871, China}
\email{fushuqi@stu.pku.edu.cn}
\affiliation{Department of Astronomy, School of Physics, Peking University, Beijing 100871, People’s Republic of China}

\author[0000-0002-2420-5022]{Zijian Zhang}
\affiliation{Kavli Institute for Astronomy and Astrophysics, Peking University, Beijing 100871, China}
\email{zjz.kiaa@stu.pku.edu.cn}
\affiliation{Department of Astronomy, School of Physics, Peking University, Beijing 100871, People’s Republic of China}

\author[0009-0003-6747-2221]{Danyang Jiang}
\affiliation{Kavli Institute for Astronomy and Astrophysics, Peking University, Beijing 100871, China}
\email{jiangdy@stu.pku.edu.cn}
\affiliation{Department of Astronomy, School of Physics, Peking University, Beijing 100871, People’s Republic of China}

\author[0000-0001-9299-5719]{Khee-Gan Lee}
\affiliation{Kavli IPMU (WPI), UTIAS, The University of Tokyo, Kashiwa, Chiba 277-8583, Japan}
\affiliation{Center for Data-Driven Discovery, Kavli IPMU (WPI), UTIAS, The University of Tokyo, Kashiwa, Chiba 277-8583, Japan}
\email{kglee@ipmu.jp}

\author[0000-0002-7633-431X]{Feige Wang}
\affiliation{Department of Astronomy, University of Michigan, 1085 S. University Ave., Ann Arbor, MI 48109, USA}
\email{fgwang@umich.edu}

\author[0000-0003-3310-0131]{Xiaohui Fan}
\affiliation{Steward Observatory, University of Arizona,933 N Cherry Ave, Tucson, AZ, 85721, USA}
\email{xiaohuidominicfan@gmail.com}

\author[0000-0001-5287-4242]{Jinyi Yang}
\affiliation{Department of Astronomy, University of Michigan, 1085 S. University Ave., Ann Arbor, MI 48109, USA}
\email{jyyangas@umich.edu}

\author[0000-0001-8496-4162]{Ruancun Li}
\affiliation{Kavli Institute for Astronomy and Astrophysics, Peking University, Beijing 100871, China}
\affiliation{Department of Astronomy, School of Physics, Peking University, Beijing 100871, People’s Republic of China}
\email{jiangdy@stu.pku.edu.cn}

\author[0000-0003-3596-8794]{Hollis B. Akins}
\affiliation{Department of Astronomy, The University of Texas at Austin, 2515 Speedway Boulevard Stop C1400, Austin, TX 78712, USA}
\affiliation{Cosmic Frontier Center, The University of Texas at Austin, Austin, TX 78712, USA}
\email{hollis.akins@gmail.com}

\author[0000-0002-1620-0897]{Fuyan Bian}
\affiliation{European Southern Observatory, Alonso de Cordova 3107, Casilla 19001, Vitacura, Santiago 19, Chile}
\email{fbian@eso.org}

\author[0000-0002-7928-416X]{Y. Sophia Dai}
\affiliation{Chinese Academy of Sciences South America Center for Astronomy, National Astronomical Observatories, CAS, Beijing 100101, China}
\email{ydai@nao.cas.cn}

\author[0000-0002-9382-9832]{Andreas L. Faisst}
\affiliation{IPAC, California Institute of Technology, 1200 E. California Blvd. Pasadena, CA 91125, USA}
\email{afaisst@caltech.edu}

\author[0000-0001-6947-5846]{Luis C. Ho}
\affiliation{Kavli Institute for Astronomy and Astrophysics, Peking University, Beijing 100871, China}
\affiliation{Department of Astronomy, School of Physics, Peking University, Beijing 100871, People’s Republic of China}
\email{lho.pku@gmail.com}

\author[0000-0001-9840-4959]{Kohei Inayoshi}
\affiliation{Kavli Institute for Astronomy and Astrophysics, Peking University, Beijing 100871, China}
\email{inayoshi.pku@gmail.com}

\author[0000-0003-0964-7188]{Linhua Jiang}
\affiliation{Kavli Institute for Astronomy and Astrophysics, Peking University, Beijing 100871, China}
\email{jiangKIAA@pku.edu.cn}
\affiliation{Department of Astronomy, School of Physics, Peking University, Beijing 100871, People’s Republic of China}

\author[0000-0002-5768-738X]{Xiangyu Jin}
\affiliation{Department of Astronomy, University of Michigan, 1085 S. University Ave., Ann Arbor, MI 48109, USA}
\email{jxiangyu@umich.edu}

\author[0000-0001-6874-1321]{Koki Kakiichi}
\affiliation{Cosmic Dawn Center (DAWN), Denmark}
\affiliation{Niels Bohr Institute, University of Copenhagen, Jagtvej 128, DK-2200, Copenhagen N, Denmark}
\email{koki.kakiichi@nbi.ku.dk}

\author[0000-0001-9187-3605]{Jeyhan S. Kartaltepe}
\affiliation{Laboratory for Multiwavelength Astrophysics, School of Physics and Astronomy, Rochester Institute of Technology, 84 Lomb Memorial Drive, Rochester, NY 14623, USA}
\email{jsksps@rit.edu}

\author[0000-0001-5951-459X]{Zihao Li}
\affiliation{Cosmic Dawn Center (DAWN), Denmark}
\affiliation{Niels Bohr Institute, University of Copenhagen, Jagtvej 128, DK-2200, Copenhagen N, Denmark}
\email{zihao.li@nbi.ku.dk}

\author[0000-0002-7214-5976]{Weizhe Liu}
\affiliation{Steward Observatory, University of Arizona,933 N Cherry Ave, Tucson, AZ, 85721, USA}
\email{oscarlwz@gmail.com}

\author[0000-0002-4544-8242]{Jan-Torge Schindler}
\affiliation{Hamburger Sternwarte, University of Hamburg, Gojenbergsweg 112, D-21029 Hamburg, Germany}
\email{jtschindler@hs.uni-hamburg.de}

\author[0000-0003-0747-1780]{Wei Leong Tee}
\affiliation{Steward Observatory, University of Arizona,933 N Cherry Ave, Tucson, AZ, 85721, USA}
\affiliation{Department of Astronomy and Astrophysics, The Pennsylvania State University, 525 Davey Lab, University Park, PA 16802, USA}
\email{wmt5159@psu.edu}


\begin{abstract}
We report the discovery of two broad-line X-ray AGNs (\cf~and \cn) at $z\sim3$ identified in the JWST Cycle 3 COSMOS-3D program using NIRCam F444W grism spectroscopy. Both exhibit prominent \hei+Pa$\gamma$ emission and absorption, indicative of circumnuclear dense gas that is traced in these systems. Complementary UV and optical spectroscopy in the COSMOS field provides Ly$\alpha$, Si {\sc iv}, and C {\sc iv} measurements. Both sources are detected in MIRI F1000W, and \cf~is also detected in F2100W, indicating hot dust emission.
 
The two AGNs show distinct black hole and obscuration properties. The source \cf~displays little red dots (LRD)-like V-shape spectra energy distribution (SED) shape with a turnover near the Balmer 4000 \AA~break, and a narrow Ly$\alpha$ line with $\log L_{\rm Ly\alpha}=42.49\pm0.01 \mathrm{erg\ s^{-1}}$, with no additional metal lines detected. In contrast, \cn~exhibits a higher \hei absorption column density, larger X-ray–inferred $N_{\rm H}$, lower intrinsic 2–10 keV luminosity, and strong blueshifted features in He {\sc i}, Si {\sc iv}, and C {\sc iv} absorption with velocity offsets exceeding $5000~\mathrm{km\ s^{-1}}$. Photoionization modeling implies gas densities of $\sim10^{9-10} \mathrm{cm^{-3}}$ and sizes comparable to the broad-line region, consistent with dense gas envelopes predicted for LRDs. Together with previous detections of \hei absorption in compact little red dots, these results suggest that dense circumnuclear gas is likely prevalent at high redshift and may regulate obscuration and black hole–host co-evolution across AGN types.
\end{abstract}


\keywords{\uat{Active galactic nuclei}{16} --- \uat{AGN host galaxies}{2017} --- \uat{Dust shells}{414} }


\section{Introduction}\label{sec:intro}

The growth of supermassive black holes and the evolution of their host galaxies are closely connected through the presence and dynamics of dense gas in the circumnuclear region. Obscuration represents a key phase in this co-evolution, where gas and dust surrounding the active nucleus attenuate the emergent radiation and generate characteristic absorption and emission signatures in the spectrum (see the reviews of \citealt{hickox2018,alex25} and references therein). However, the physical processes that govern the structure and evolution of the obscuring gas, and how this gas regulates both black hole accretion and star formation in the host galaxy, remain poorly understood.


Theoretical models suggest that obscuration is closely tied to luminosity-dependent AGN demographics \citep{gilli07,lizijian2024}, with circumnuclear hydrogen volume densities reaching $n_{\rm H}\gtrsim10^{8},\mathrm{cm^{-3}}$ \citep{1995ApJ...455L.119B, wuqiaoya25}. Observationally, X-ray surveys have established that obscuration is common among AGNs across cosmic time. In the COSMOS field, $\sim$60\% of X-ray AGNs are obscured \citep{marchesi16}, while local Swift--BAT observations indicate an obscured fraction of $\sim$70\% \citep{2017ApJ...850...74K, cricci17}. Deep Chandra observations further reveal that the obscured fraction increases with redshift and includes a substantial Compton-thick population \citep{buchner15, liu17}, with hard X-ray observations from NuSTAR confirming the prevalence of heavily obscured AGNs \citep{alex13,arevalo17,peca25}.

\hei absorption offers distinct advantages for probing the physical conditions of the circumnuclear absorbing gas. The \hei~transition arises from recombination of He$^{+}$ to the metastable $2^{3}$S level (ionization potential of 24.56 eV). Because the diffuse stellar radiation field is weak above 24.56 eV \citep{ji15}, the presence of \hei absorption is a robust indicator of AGN ionization. Moreover, its large oscillator strength ($f=0.5392$) provides a wider dynamic range in column density compared to commonly used UV resonant lines such as C~{\sc iv} and Si~{\sc iv} \citep{leighly14}. However, because \hei lies outside the typical optical spectral window, reported detections remain limited. \citet{liu15} compiled only 11 quasars with \hei absorption known at the time, most at $z<1$. The first data release of NIR spectroscopy from the BASS survey identified 7 out of 102 local X-ray AGNs with possible \hei absorption, and the number continues to grow in later releases (see, e.g., Figure~2 in \citealt{ricci22}).

The advent of JWST has enabled higher sensitivity and broader wavelength coverage in the near- and mid-infrared. One of the most intriguing discoveries from JWST is a population of compact, red sources known as ``little red dots'' (LRDs), which exhibit a characteristic V-shaped spectral energy distribution, with rest-frame UV continua that are bluer than the optical ones and a turnover near the Balmer break. The $z>4$ LRDs are frequently detected with broad Balmer emission lines \citep[e.g.,][]{matthee24,kocevski25}, indicative of AGN-like activity. Their physical nature is still under debate, with the main discussion focusing on the relative contribution from host galaxies versus AGN emission. Several studies interpret LRDs as dust-reddened AGNs at early cosmic times \citep[e.g.,][]{harikane23, kocevski23, maiolino23, matthee24} or stellar light from its host galaxy \citep[e.g.,][]{greene24}. Another scenario is that LRDs are AGN embedded in dense gas envelopes \citep{kohei25,degraaff25}, which is supported by LRD observations with Balmer absorption lines at $z>4$ and a small number of \hei\ absorption at $z\sim3$ \citep{juodzbalis24, wang25, naidu25,kokorev25,loiacono25,graff25}. 
Interestingly, \citet{lin25} identified three local analogs of LRDs in SDSS, all of which also show \hei~absorption. 

Most LRDs are found to be both X-ray and radio weak \citep{yue24}. Recently, two sources reported by \citet{shuqi25} (with Forge-II corresponding to \cf~in this work) were identified as LRDs with V-shape SEDs and compact morphologies and interpreted as transitioning into quasars. In addition, an “X-ray dot” reported by \citet{xraydot} exhibits Balmer absorption, an LRD-like SED, and a $2$--$10~$ keV X-ray luminosity of $10^{44.18}~\mathrm{erg~s^{-1}}$, but is mid-infrared faint at $z\sim3$. These observations suggest that linking gas phases across different stages of AGN growth is a promising approach to understanding the role of obscuration in black hole growth and host galaxy evolution at high redshift. Motivated by this, comparing X-ray–bright, LRD-like AGNs with the broader AGN population is essential for clarifying the nature of LRDs, high-redshift red quasars, and the role of gas and dust in their growth.

In this letter, we report the detections of two AGNs at $z\sim3$ in the COSMOS field with both X-ray and MIRI observations. Both sources exhibit broad \hei\ emission (full width at half maximum, FWHM $>$ 2000 km s$^{-1}$) and prominent absorption features, indicating the presence of dense gas surrounding the broad-line region. The \hei\ spectra and MIRI data are drawn from the JWST GO~3 large program COSMOS-3D (hereafter C3D, PID \#5893; PI: K.~Kakiichi), which provides F444W NIRCam grism spectroscopy along with MIRI F1000W and F2100W imaging. The data come from observations conducted between December 2024 and May 2025. An overview of the program and its major science goals will be presented in Kakiichi et al., in prep. In Section~\ref{sec:target_selection}, we describe the target selection and ancillary datasets. Section~\ref{sec:proper} presents measurements of the AGN properties, including stellar mass ($M_*$), black hole mass ($M_{\rm BH}$), and star formation rate (SFR). In Section~\ref{sec:discussion}, we discuss the role of dense gas traced by \hei\ absorption in obscuration and black hole growth. Section~\ref{sec:cloudy} models the photoionization state of this gas, and Section~\ref{sec:dndz} compares these two AGNs with other populations, including LRDs.

Throughout the paper, we adopt a standard $\Lambda$CDM cosmology with $H_0 = 70\ {\rm km}\ {\rm s}^{-1}\ {\rm Mpc}^{-1}$, $\Omega_{M}$ = 0.3 and $\Omega_{\lambda}$ = 0.7. 
The initial mass function (IMF) adopted in this paper is from \citet{chabrier}.


\section{Target Selection and Observations} \label{sec:target_selection}

\subsection{Target selection}
The two AGNs were first selected from the X-ray catalog of the Chandra COSMOS-Legacy (CCL) Survey \citep{civano16}, which is a 4.6~Ms \textit{Chandra} program covering 2.2~deg$^{2}$ of the COSMOS field. \citet{marchesi16b} reported X-ray properties for 1,855 sources with $>$30 net counts and derived intrinsic hydrogen absorption column density ($N_{\rm H}$) and rest-frame 2 to 10~keV luminosities ($L_{\rm X,2-10 keV}$). Among the 1,238 X-ray AGNs (log~$L_{\rm X,2-10 keV}  > 42$, AGN selection criteria  from \citealt{alexander05}) with secure spectroscopic redshifts, 342 fall within the C3D NIRCam imaging footprint, and 106 of these 342 are covered by C3D MIRI observations (either 1000W or 2100W). We visually inspected the F444W grism spectra of these 106 AGNs, and two of them, \cf~and \cn, show prominent \hei~absorption features. The information and measured properties of these two AGNs described in Section \ref{sec:proper} are listed in Table~\ref{table:prop}. We refer to \citet{shuqi25}, who report \cf~and another X-ray AGN as two LRDs transitioning into quasars at $z\sim3$, providing detailed rest-frame UV data and discussion. In this work, we focus on investigating the roles of absorbing gas and hot dust at different stages of black hole growth.


\subsection{JWST NIRCam and MIRI imaging}

These two targets are covered by both C3D and the GO1 treasury program COSMOS-Web (PI: C. Casey, PID~\#1727). In C3D, we conducted NIRCam grism spectroscopy using F444W filter, with the F200W filter simultaneously employed in the short-wavelength (SW) channel during the WFSS observations. Direct imaging was obtained with the F115W and F356W filters, providing a total NIRCam coverage of $\sim1151~{\rm arcmin^2}$. COSMOS-Web have NIRCam imaging in four filters (F115W, F150W, F277W, F444W) and MIRI F770W in parallel. We obtained the COSMOS-Web data from their DR1 data and catalog release 1 \citep{shuntov25}\footnote{https://cosmos2025.iap.fr/}. All images (both NIRCam and MIRI) are drizzled to a $0.03^{\prime\prime}$ pixel scale. For C3D, we performed the NIRCam and MIRI imaging reduction on F115W, F200W and F356W bands using the JWST pipeline version 1.16.1 \citep{jwst_pipeline}, with the CRDS calibration reference file context \texttt{jwst\_1303.pmap}. 

The C3D MIRI data (F1000W and F2100W) were processed following the approach adopted in the Systematic Mid-infrared Instrument Legacy Extragalactic Survey (SMILES; \citealt{Lyu24}). During Stage 2 reduction, we implemented a customized external background subtraction module from the Rainbow Database JWST pipeline, which constructs a super-background model to suppress cosmic-ray–induced artifacts \citep{alvarez23}.

\subsection{JWST Grism spectra}

The WFSS spectra were taken in the F444W filter with Grism R. The $5\sigma$ line flux limit is $4\times10^{-18}~{\rm erg~s^{-1}~cm^{-2}}$ at $\sim4.5~\mu{\rm m}$ (i.e., $m_{\rm F444W}\lesssim27.5$) within the C3D NIRCam coverage. The spectral resolution is $R\sim1600$ at $\sim4~\mu$m. We reduced the grism data following the method described in \citet{sun23}\footnote{\url{https://github.com/fengwusun/nircam_grism}}. We briefly describe the procedure as follows: the Stage 1 data were processed with the standard JWST pipeline. The data were flat fielded using imaging flats taken with the same filter and module, and the sky background was removed using sigma-clipped median grism frames. The wavelength and spectral tracing solutions were derived from commissioning calibrations and refined using field dependent grism trace and dispersion models. Flux calibration was performed using spectra of the standard star P330-E, and the calibrated 2D spectra were extracted and combined for all F444W grism exposures. 

\begin{table}[h]
\centering
\caption{Physical properties of the two AGNs. Stellar masses and SFRs are derived with \texttt{CIGALE}. The outflow velocity $v_{max}$ is taken from \citet{science}.}
\begin{tabular*}{\columnwidth}{@{\extracolsep{\fill}}ll}
\hline
\multicolumn{2}{c}{\bf \cf~} \\ 
\hline
RA, DEC & 149.89197, 2.28509\\
Redshift & 2.933\\
log L$_{\rm X,2-10\ keV}$ (erg s$^{-1}$) & 45.17\\
log L$_{\rm bol}$ (erg s$^{-1}$) & 46.69\\
log $N_{\rm H}$ (cm$^{-2}$) & 22.50$^{+0.33}_{-1.16}$\\
log M$_{\ast}$/M$_{\odot}$ & 10.92$^{+0.29}_{-0.29}$\\
log SFR (M$_{\odot}$ yr$^{-1}$) & 2.55$^{+0.19}_{-0.19}$\\
log M$_{\rm BH}$/M$_{\odot}$ & 9.15$^{+0.4}_{-0.4}$\\
$\lambda_{\rm Edd}$ & 0.28$_{-0.18}^{+0.41}$\\
$f_{\mathrm{HeI\lambda10830}}$ emission (10$^{-17}$ erg s$^{-1}$ cm$^{-2}$) & 5.94$^{+0.15}_{-0.15}$\\
FWHM$_{\mathrm{HeI\lambda10830}}$ broad (km s$^{-1}$) & 3591$^{+78}_{-78}$\\
FWHM$_{\mathrm{HeI\lambda10830}}$ narrow (km s$^{-1}$) &
1193$^{+32}_{-32}$\\
EW$_{\mathrm{HeI\lambda10830}}$ emission broad (\AA) & 
135$^{+2}_{-2}$\\
EW$_{\mathrm{HeI\lambda10830}}$ emission narrow (\AA) & 
16$^{+1}_{-1}$\\
$f_{\rm Pa\gamma}$ emission (10$^{-17}$ erg s$^{-1}$ cm$^{-2}$) & 1.66$^{+0.04}_{-0.04}$\\
FWHM$_{\rm Pa\gamma}$ broad (km s$^{-1}$) & 2553$^{+55}_{-55}$\\
FWHM$_{\rm Pa\gamma}$ narrow (km s$^{-1}$) & 1195$^{+33}_{-33}$\\
EW$_{\rm Pa\gamma}$ emission broad (\AA) & 42$^{+1}_{-1}$\\
EW$_{\rm Pa\gamma}$ emission narrow (\AA)& 1$^{+1}_{-1}$ \\
EW$_{\mathrm{HeI\lambda10830}}$ absorption (\AA) & 31$^{+6}_{-6}$\\
log$N$(\hei) (cm$^{-2}$) & 13.9$^{+0.1}_{-0.1}$\\
\hline
\multicolumn{2}{c}{\bf \cn~} \\ 
\hline
RA, DEC & 150.29725, 2.14884\\
Redshift & 3.328\\
log L$_{\rm X,2-10\ keV}$ (erg s$^{-1}$) & 44.62\\
log L$_{\rm bol}$ (erg s$^{-1}$) & 46\\
log $N_{\rm H}$ (cm$^{-2}$) & 23.47$^{+0.29}_{-0.47}$\\
log M$_{\ast}$/M$_{\odot}$ & 11.01$^{+0.19}_{-0.20}$\\
log SFR (M$_{\odot}$ yr$^{-1}$) & 1.49$^{+0.20}_{-0.20}$\\
log M$_{\rm BH}$/M$_{\odot}$ & 9.84$^{+0.28}_{-0.28}$\\
$\lambda_{\rm Edd}$ & 0.013$_{-0.006}^{+0.006}$\\
$f_{\mathrm{HeI\lambda10830}}$ emission (10$^{-17}$ erg s$^{-1}$ cm$^{-2}$) & 1.90$^{+0.06}_{-0.06}$\\
FWHM$_{\mathrm{HeI\lambda10830}}$ broad (km s$^{-1}$) & 2473$^{+49}_{-49}$\\
FWHM$_{\mathrm{HeI\lambda10830}}$ narrow (km s$^{-1}$) & 1198$^{+38}_{-38}$\\
EW$_{\mathrm{HeI\lambda10830}}$ emission broad (\AA) & 75$^{+2}_{-2}$\\
EW$_{\mathrm{HeI\lambda10830}}$ emission narrow (\AA) &  5$^{+1}_{-1}$\\
$f_{\rm Pa\gamma}$ emission (10$^{-17}$ erg s$^{-1}$ cm$^{-2}$) & 0.67$^{+0.02}_{-0.02}$\\
FWHM$_{\rm Pa\gamma}$ broad (km s$^{-1}$) & 4353$^{+140}_{-140}$ \\
FWHM$_{\rm Pa\gamma}$ narrow (km s$^{-1}$) &  1196$^{+38}_{-38}$\\
EW$_{\rm Pa\gamma}$ emission broad (\AA) & 24$^{+1}_{-1}$\\
EW$_{\rm Pa\gamma}$ emission narrow (\AA) & 6$^{+1}_{-1}$\\
EW$_{\mathrm{HeI\lambda10830}}$ absorption (\AA) & 33$^{+10}_{-10}$\\
log$N$(\hei) (cm$^{-2}$) & 14.1$^{+0.2}_{-0.2}$\\
$v_{max}$ (km s$^{-1}$) & $\sim$12000\\
\hline
\end{tabular*}
\label{table:prop}
\end{table}

\subsection{Ancillary data}\label{ancillary_data}

The COSMOS field provides extensive multiwavelength ancillary observations, allowing a detailed characterization of these two targets from X-ray to radio. Below we summarize the datasets used in this work.

\subsubsection{X-ray}
We use the reduced archival X-ray data from the CCL survey and its X-ray catalog \citep{marchesi16b} in this work (see Figure \ref{fig:xray_data}). CCL survey has a hard-band 2-10 keV flux limit of $1.5\times10^{-15}~\mathrm{erg~s^{-1}~cm^{-2}}$. The data were reduced with CIAO 4.5 \citep{ciao} and CALDB 4.5.9. The intrinsic 2–10 keV X-ray luminosity $L_{2\text{–}10,\mathrm{keV}}$ and the absorbing column density $N_{\rm H}$ are derived from spectral fits using an absorbed power–law model and the photon index $\Gamma$ is fixed at 1.9 in the fitting. 
Details of the X-ray spectral fitting are provided in \citet{marchesi16b}. The $L_{2-10~\mathrm{keV}}$ of \cf\ is $L_{2-10~\mathrm{keV}}=10^{45.17}~\mathrm{erg~s^{-1}}$ with a hydrogen column density $N_{\mathrm{H}}=10^{22.5}~\mathrm{cm^{-2}}$. For \cn, the $L_{2-10~\mathrm{keV}}$ is $10^{44.62}~\mathrm{erg~s^{-1}}$ and the column density is $N_{\mathrm{H}}=10^{23.47}~\mathrm{cm^{-2}}$. Both sources are therefore classified as Compton-thin ($N_{\mathrm{H}}<10^{24}~\mathrm{cm^{-2}}$) obscured AGN based on their hard X-ray detections and X-ray inferred hydrogen column densities.


\subsubsection{UV--mid-IR}

For the bands not covered by the JWST imaging data, we collect the UV-to-mid-IR photometry using the Kron AUTO measurements from the COSMOS2020 Classic catalog \texttt{COSMOS2020} \citep{cosmos2020}, spanning from the \textit{GALEX}/FUV band to the \textit{Spitzer}/IRAC channel 4. 
The $u$ band photometry is from the CFHT Large Area U band Deep Survey (CLAUDS) program \citep{clauds}.
The ongoing HSC Subaru Strategic Program (HSC-SSP) survey provides imaging from $g$ band to $y$ band \citep{hscssp}. 
Besides, we include the HST/F814W photometry from \citet{hst814}. 
The HST/F814W AB magnitudes are 24.49$\pm$0.05 and 20.50$\pm$0.01 for \cf~ and \cn, respectively. 
The $YJHK_{s}$ data is from the fourth data release  of the UltraVISTA survey \citep{ultravista}.
For mid-IR photometry, we use the magnitudes from the
Spitzer Large Area Survey with Hyper Suprime-Cam
(SPLASH; \citealt{splash}).
Neither of the two objects are detected in the $GALEX$/FUV or NUV bands.


\subsubsection{submm--FIR--radio}

The \cf\ has been observed in multiple ALMA programs, including \#2016.1.01001.S (PI: J. Kartaltepe; 887 s integration time in Band 3), \#2021.1.01328.S (PI: W. Rujopakarn; 60 s in Band 7), \#2021.1.00225.S (PI: C. Casey; 45 s in Band 4), and \#2023.1.00180.L (CHAMPS; PI: A. Faisst; Faisst et al., in prep.; 54 s in Band 6). There is no $>3\sigma$ dust continuum detection in the Band 3, 6 and 7 observation for \cf. The \cn\ has a reliable (S/N $\sim$ 7.5) ALMA Band 6 (1.1 mm) detection from program \#2016.1.01012.S (PI E. Treister). We retrieved the reduced data from the latest public release of the A3COSMOS program \citep{a3cosmos}.

We match the FIR photometry of these two sources using the super-deblended catalog \citep{superdeblended}, which combines FIR to (sub)millimeter measurements in the COSMOS field. The Herschel/PACS 100 and 160 $\mu$m data are from the PEP program (PI D. Lutz; \citealt{pep}), and the SPIRE 250, 350, and 500 $\mu$m data are from the Herschel Multi-tiered Extragalactic Survey (PI: S. Oliver; \citealt{hermes}). Since the super-deblended catalog is based on NIR prior positions, we adopt a matching radius of 1 arcsec. Neither source has S/N $>$ 3 detections at 100 or 160 $\mu$m.

At radio wavelengths, we search the VLA-COSMOS 1.4 GHz \citep{vla1.4ghz} and 3 GHz \citep{vla3ghz} catalogs. The \cf\ is detected with high significance (S/N $>$ 10) at both 1.4 GHz and 3 GHz, while \cn\ shows no radio detection.

\subsubsection{Ground-based spectroscopy}
We find that \cf~has archival Keck/LRIS (exptime = 2h, $R\sim$ 1100) spectroscopy from the COSMOS Lyman-Alpha Mapping and Tomography Observations (CLAMATO) survey \citep{clamato,clamato2}, covering 3200–5500~\AA, and Keck/DEIMOS (exptime = 1800s) spectroscopy from the DEIMOS 10k survey \citep{deimos}, covering 5500–9800 \AA~and spectral resolution $R\sim$ 2000. The two parts of the spectra are shown in the Appendix Figure \ref{fig:optical_spec}. A clear \lya~emission line is detected at $z = 2.933$ with a signal-to-noise ratio of $\sim45$, while no C~{\sc iv} or C~{\sc iii}] lines are detected. This is further confirmed by DESI spectra presented in \citet{shuqi25}. The \lya~line is well fitted by a Gaussian profile with a FWHM of 744.90$\pm$32.52~km~s$^{-1}$. The apparent \lya~luminosity is 10$^{42.49}$ erg s$^{-1}$. 
This value is similar to that detected in a sample of 23 $z\sim2$ type~II AGN selected from SDSS, part of which show both narrow \lya~and broad Balmer lines \citep{wangben25}.

The \cn\ was observed in the zCOSMOS-bright survey \citep{zcosmos}, showing significant ($\sim12000$ km s$^{-1}$) blueshifted outflow features in the Si~{\sc iv}, C~{\sc iv}, and Al~{\sc iii} absorption lines (see the right panel of Figure \ref{fig:optical_spec}). \citet{science} also obtained a $K$-band (1.9–2.4 $\mu$m) spectrum of this target with Keck/MOSFIRE. The spectrum shows a broad H$\beta$ line with FWHM $=11330$ km s$^{-1}$, yielding a black hole mass estimate of $6.91\times10^{9}\ M_{\odot}$ from the broad-line virial method.

\section{Measurements and Properties}\label{sec:proper}

\begin{figure*}[ht!]
\epsscale{1.15}
\plottwo{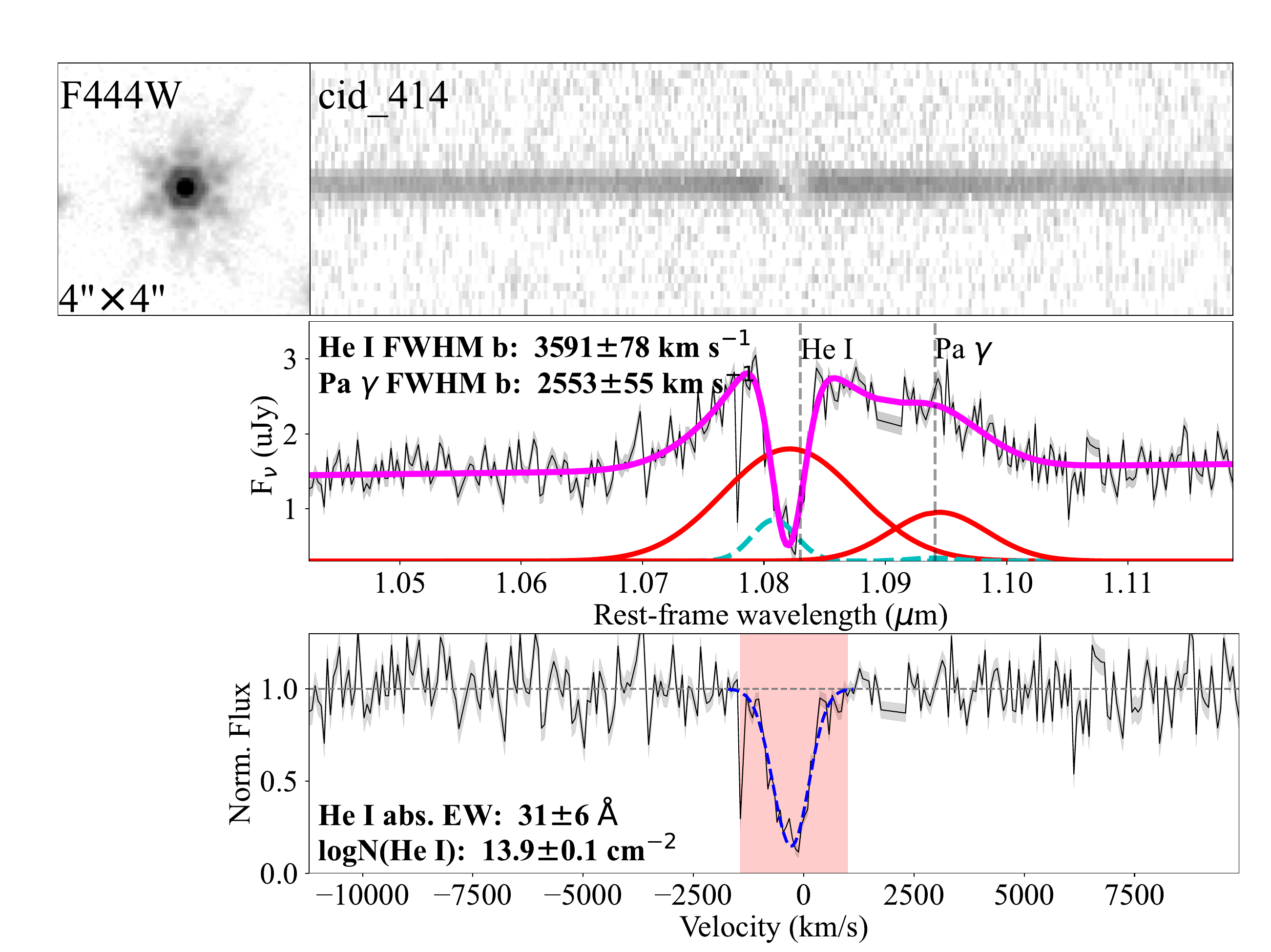}{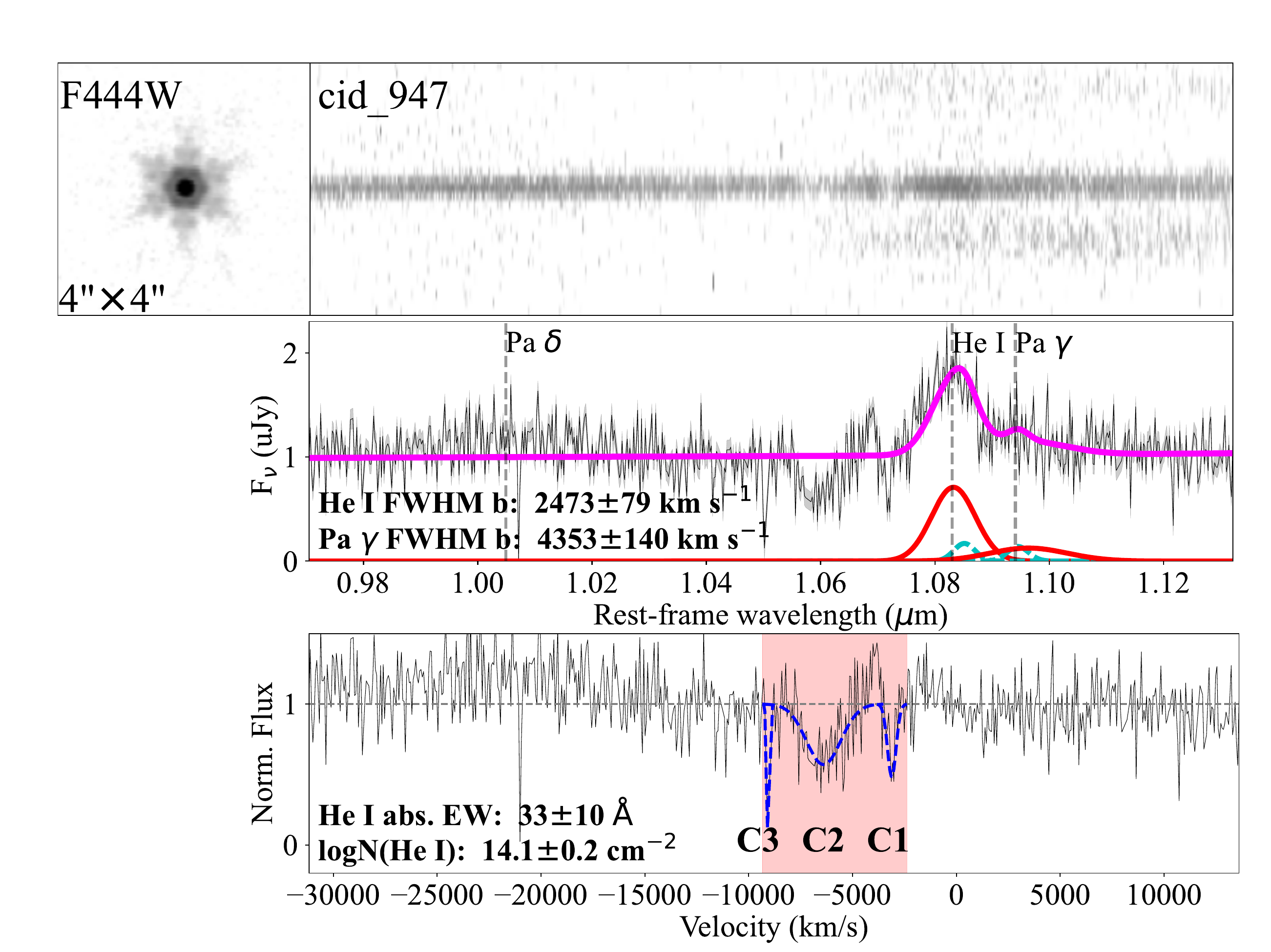}
\caption{The JWST 2D, 1D grism F444W spectra and normalized  spectra of \cf~(left) and \cn~(right) from C3D. The Gaussian fits to the \hei–\pag\ complex are shown as red curves in the middle panels. The broad and narrow components are plotted as red solid lines and cyan dashed curves, respectively. In the bottom panels, the absorption profiles are plotted with blue dashed lines. For \cf~, there is one absorption component while there are three components in the spectrum of \cn~ (marked as C1, C2 and C3 in the right bottom panel). 
}\label{fig:spec}
\end{figure*}

\subsection{Emission and absorption line measurements}\label{sec:emi_abs}

We present the NIRCam F444W imaging, and the 2D (without conitinuum subtraction) and 1D grism spectra of \cf~and \cn~in Figure~\ref{fig:spec}. For the two targets, we first fit the He~{\sc i}+Pa$\gamma$ line structure using Gaussian profiles for both the emission and absorption components. A linear continuum is fitted around the emission lines in the rest-frame wavelength ranges 1.02–1.05~$\mu$m and 1.11–1.14~$\mu$m. We fit the \hei~and Pa$\gamma$ emission lines using three models: broad only, narrow only, and broad+narrow components, and we find that the resulting $\chi^{2}$ values are similar. In Figure~\ref{fig:spec}, we present the fitting results for the two targets using a model with one broad and one narrow component. We describe the details of the emission and absorption line measurements of the two targets below and in Table \ref{table:prop}.
\subsubsection{\cf}

For \cf, we fix the center of the \hei~and Pa$\gamma$ lines to the redshift determined from the \lya~line ($z = 2.933$, vertical dashed line in Figure~\ref{fig:spec}). The FWHM of the broad (narrow) component of \cf~is 3591$\pm$78 (1193$\pm$32) km s$^{-1}$ for the \hei~ line. The line widths for the \pag are 2553$\pm$55 and 1195$\pm$33 km s$^{-1}$ for the broad and narrow components, respectively. We correct the line instrumental broadening based on the NIRCam grism line-spread function \citep{greene17}. The equivalent widths (EWs) of the \hei~and Pa$\gamma$ broad emission lines are 135$\pm$2~\AA~and 16$\pm$1~\AA, respectively.

We use the apparent optical depth (AOD) method to estimate the column density of the \hei~absorption line. The optical depth is defined as $\tau(v) = -\ln \left( I(v)/I_{\rm c}(v) \right)$, where $I(v)$ and $I_{\rm c}(v)$ are the observed spectral intensity and the interpolated absorption–free continuum, respectively. The column density is then calculated by integrating the optical depth over the velocity range of the absorption:
\begin{equation}
N = \frac{m_{e} c}{\pi e^{2} \lambda f} \int_{v_{\rm min}}^{v_{\rm max}} \tau(v)~dv ,
\end{equation}
where $m_{e}$ and $e$ are the electron mass and charge, $c$ is the speed of light, $\lambda$ is the rest-frame transition wavelength, and $f$ is the oscillator strength. We obtain for \cf~a column density of log~$N$(\hei)/cm$^{-2}$ = 13.9 $\pm$ 0.1.


\subsubsection{\cn~}
Different from \cf, where the \hei\ absorption does not show a significant velocity offset from the emission-line center, the object \cn\ is a BAL quasar exhibiting a significant velocity offset ($> 5000$ km s$^{-1}$) in the absorption features in both the rest-frame UV lines and \hei. The fluxes of \hei~and \pag~emission lines for \cn\ are $(1.90 \pm 0.06) \times 10^{-17}$ and $(0.67 \pm 0.02) \times 10^{-17}$ erg s$^{-1}$ cm$^{-2}$, respectively. From the right panel of Figure \ref{fig:optical_spec}, we see that the \hei\ line profile shares a similar velocity width with the rest-frame UV absorption lines. We fit the \hei\ absorption with three major components  (as C1,2 and 3) in Figure \ref{fig:spec}, and list the corresponding measured equivalent widths  and the total column density in Table \ref{table:prop}. There is likely a fourth subcomponent at a velocity offset of $\Delta v\sim -12{,}000$ km s$^{-1}$, but since it is not detected above $3\sigma$ in the 2D spectra, we do not include it when calculating the \hei~absorption-line column density. The total \hei~absorption-line column density, integrated over the three subcomponents shown in Figure~\ref{fig:spec}, is 10$^{14.1 \pm 0.2}$~$\mathrm{cm^{-2}}$.

\subsection{JWST photometry}\label{sec:phot}



For all JWST NIRCam and MIRI bands used in this study, we performed photometric measurements using SEP for source detection and SE++ for photometry \citep{bertin20}. SEP was applied to the square root of a positive $\chi^2$ detection image constructed from all available broadband images within each tile. It was run in both “hot” and “cold” modes, which were later merged to optimize the detection of bright, extended galaxies and deblended faint, compact sources. Automatic Kron photometry was carried out using “small Kron” (k=1.2, R=1.6) and “default Kron” (k=2.5, R=3.5) apertures with SEP and \texttt{photutils}, and the ratio of these fluxes was used for aperture correction. The images are not point-spread-function (PSF)-matched; however, the Kron shape parameters are derived from the $\chi^2$ image. SE++ was subsequently run in ASSOC mode to compute fixed-aperture photometry for the SEP-detected sources. The complete C3D photometry details and catalog will be described in Champagne et al., in prep. 


\begin{figure*}[ht!]
\epsscale{1.15}
\plottwo{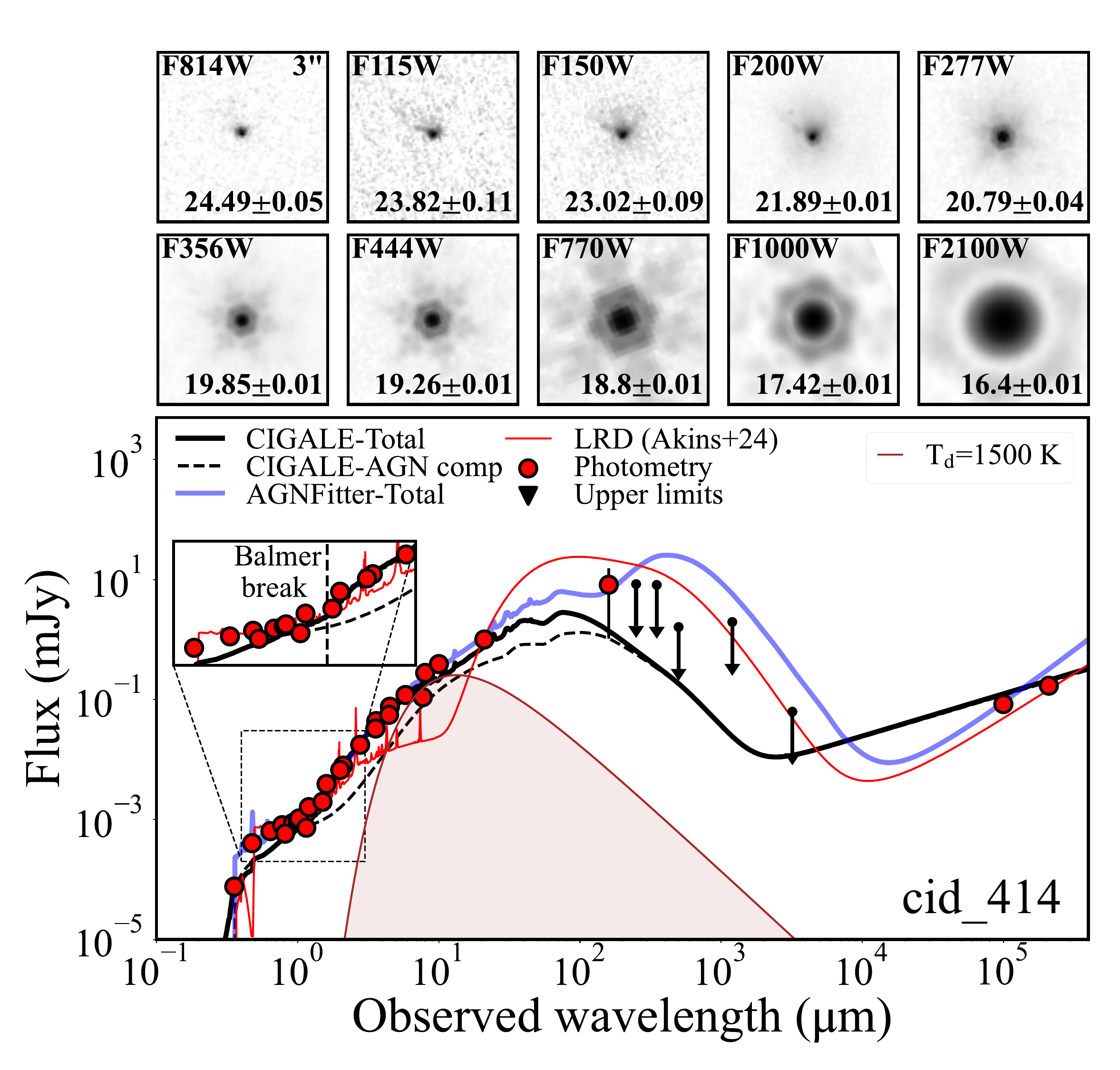}{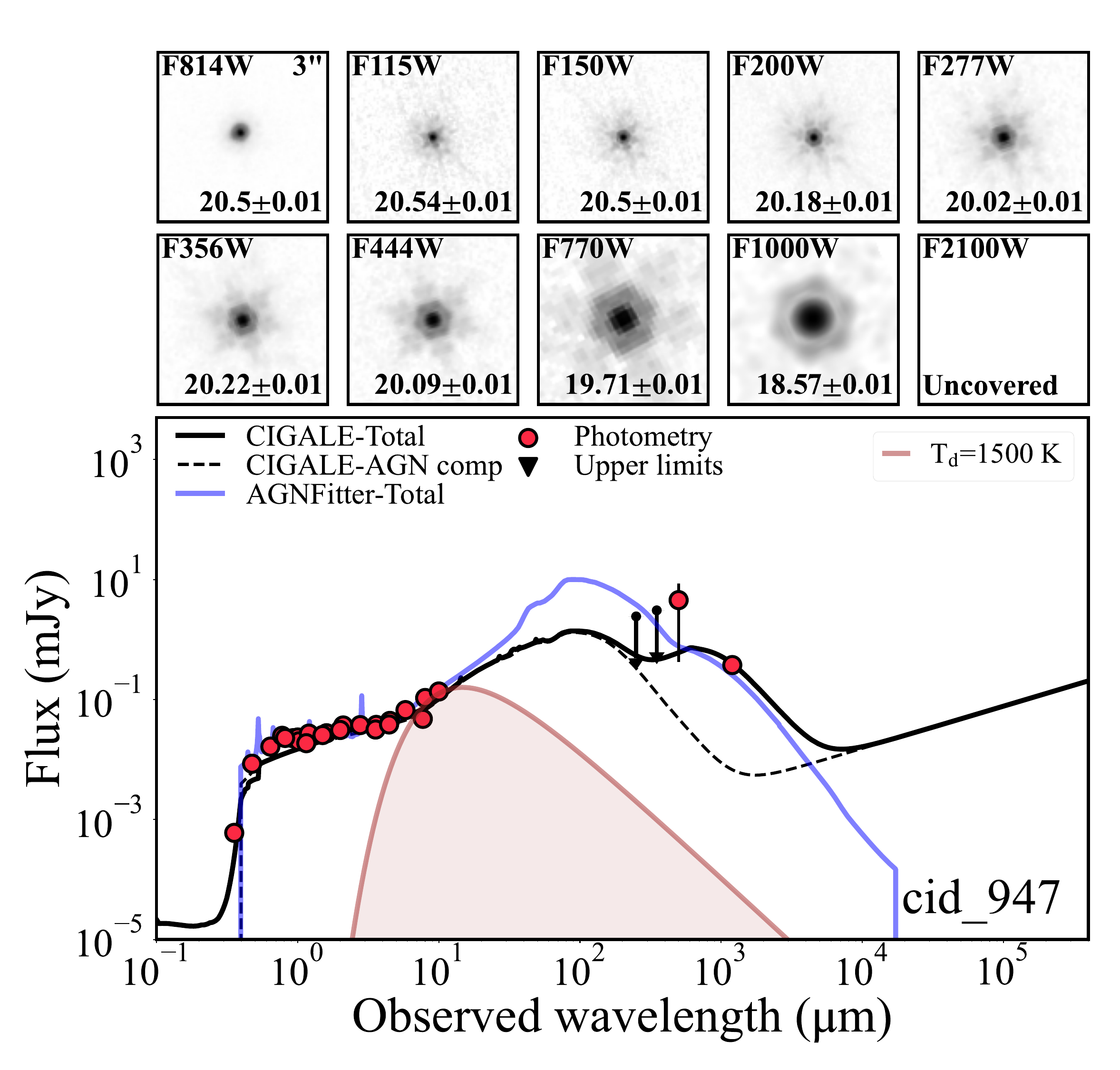}
\caption{Multiwavelength images and SED fitting results of \texttt{cid\_414} (left) and \texttt{cid\_947} (right). We present the SED fitting from \texttt{CIGALE} (black) and \texttt{AGNFittter} (blue). The dashed curves represent the best-fit AGN emission in \texttt{CIGALE}. Red dots with error bars indicate the observed photometry, while black arrows show the upper limits. We add a zoom-in window for \cf~to present the similarity with LRD's V-shape turnover around the Balmer break.}
\label{fig:sed}
\end{figure*}

\subsection{Point-Spread Function Construction and Image Decomposition}\label{sec:decomposition}

In order to measure the host galaxy (if present) properties, we perform careful imaging decomposition. Previous studies usually adopt two components to model the light distribution: a PSF profile for the AGN and a S\'ersic profile for the host galaxy \citep{ding20, li2021,zhuang24b}. The PSF is constructed from stars within a $2^{\prime\prime}$ radius around the targets. We use 11 stars for \cf\ and 5 for \cn. The stellar cutouts are oversampled by a factor of three, recentered, and median stacked to generate the final PSF.

We use \texttt{SExtractor} to identify stellar candidates in each tile and band, excluding sources near image edges or affected by bad pixels. From the remaining objects, we select stars with consistent half-light radii and extract $6\arcsec$ cutouts, within which $\sim98\%$ of the stellar light is enclosed \citep{zhuang24}. The \texttt{SExtractor} segmentation map masks unrelated sources.

We then perform the image decomposition using the code \texttt{GALFITS} \citep{galfits}. Different from traditional single-band decomposition tools, \texttt{GALFITS} simultaneously fits images taken in multiple bands and incorporates physically motivated SED models into the fitting process. This enables the simultaneous extraction of the physical properties of both the AGN and the host galaxy. As shown in Figure~\ref{fig:galfits}, \cn\ appears morphologically consistent with a point source, while \cf\ exhibits asymmetric structures in the JWST F115W, F150W, and F200W bands. About $\sim51\%$ of the total flux is enclosed within a $0\farcs18$ aperture in F200W ($82\%$ for \cn). The F200W emission extends to $\sim7.9$~pkpc to the north. This extension may trace H$\beta$+[O~{\sc iii}] emission in F200W and Mg~{\sc ii}+[O~{\sc ii}] emission in F115W/F150W \citep{shuqi25}. For \cn, a single S\'ersic plus PSF model yields a stellar mass of $10^{10.94}\ M_{\odot}$. We find that if we include the extended region and fit \cf~with a single S\'ersic component, the derived stellar mass is $10^{11.83}\ M_{\odot}$, comparable to that of a very massive quiescent galaxy in the local universe. We therefore adopt two S\'ersic components to model the host galaxy emission of \cf, without imposing constraints on the S\'ersic index or effective radius because of its asymmetric morphology. 
We exclude the flux from the second S\'ersic component when estimating the host galaxy mass and in the subsequent SED fitting in Section~\ref{sec:sed}. The stellar mass derived from the primary S\'ersic component is $10^{11.12\pm0.33}\ M_{\odot}$. For a detailed decomposition of the two components, we refer to \citet{shuqi25}, where the second component is interpreted as nebular emission.

\subsection{SED fitting and host galaxy properties}\label{sec:sed}

To measure the host galaxy and black hole properties of the two targets, we perform multiwavelength spectral energy distribution (SED) fitting using UV–to–radio photometry introduced in Section \ref{sec:target_selection}. To mitigate systematics arising from the assumptions built into any single SED fitting tool (choices of AGN and stellar components, dust treatment, and star formation history etc), we use two independent codes, \texttt{CIGALE} \citep{cigale} and \texttt{AGNfitter} \citep{agnfitter}. \texttt{CIGALE} fits the SED with energy balance, adopting BC03 stellar models \citep{bc03}, a Chabrier IMF \citep{chabrier}, a delayed-exponential star formation history (SFH), the \citet{calzetti00} attenuation law, \citet{draine14} dust emission, and the SKIRTOR AGN model \citep{skirtor1,skirtor2}. \texttt{AGNfitter} uses the same stellar and attenuation models but does not enforce energy balance; it models dust emission with \citet{schreiber18} templates and decomposes AGN emission into an accretion-disk big blue bump (BBB) component \citep{temple21} and a dusty torus described by the CAT3D library \citep{cat3dwind}.

The best-fit SEDs from \texttt{CIGALE} and \texttt{AGNfitter} 
are shown in black and blue in Figure \ref{fig:sed}, respectively.  We note that the UV optical SED of \cf\ shows a turnover around the Balmer break (see the zoom in window in the left panel together with the LRD stacking SED template from \citealt{akins}). Combined with its compact morphology and the absence of UV metal lines, \cn\ exhibits LRD like features, except for the presence of X ray and MIRI detections. In the \texttt{AGNfitter} results, the rest-frame UV to mid-IR emission of both targets is dominated by the AGN. In the \texttt{CIGALE} results, \cn\ is AGN-dominated, while \cf\ shows a non-negligible contribution from stellar emission. In particular, if we include the fluxes from both S\'ersic components for \cf, the derived stellar masses are $10^{12.09}\ M_{\odot}$ and $10^{12.08}\ M_{\odot}$ from \texttt{CIGALE} and \texttt{AGNfitter}, respectively, which are unrealistically high for a typical star-forming galaxy at $z\sim3$. We therefore adopt only the S\'ersic 1 component in the SED fitting, resulting in a stellar mass of $\sim10^{11.12\pm0.33}M_{\odot}$. We further use the F200W (rest frame $V$ band) S\'ersic 1 flux and the mass–to–light ratio from \citet{faber79}, varying the $M/L$ from 2 to 7.6 as a consistency check, yielding a stellar mass of $10^{10.92\pm0.29}M_{\odot}$, as listed in Table~\ref{table:prop}.

\begin{figure*}[ht!]
\epsscale{1.15}
\plottwo{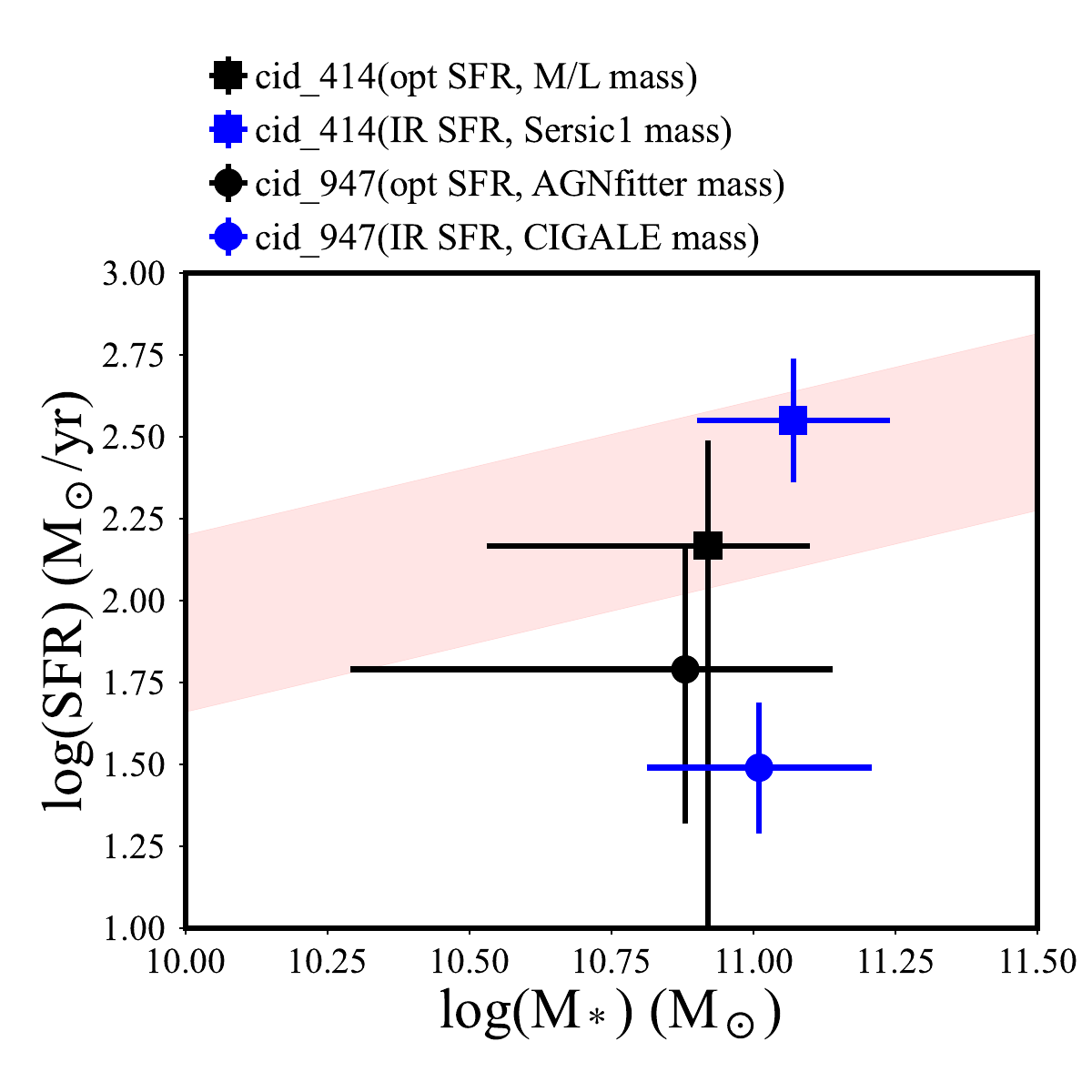}{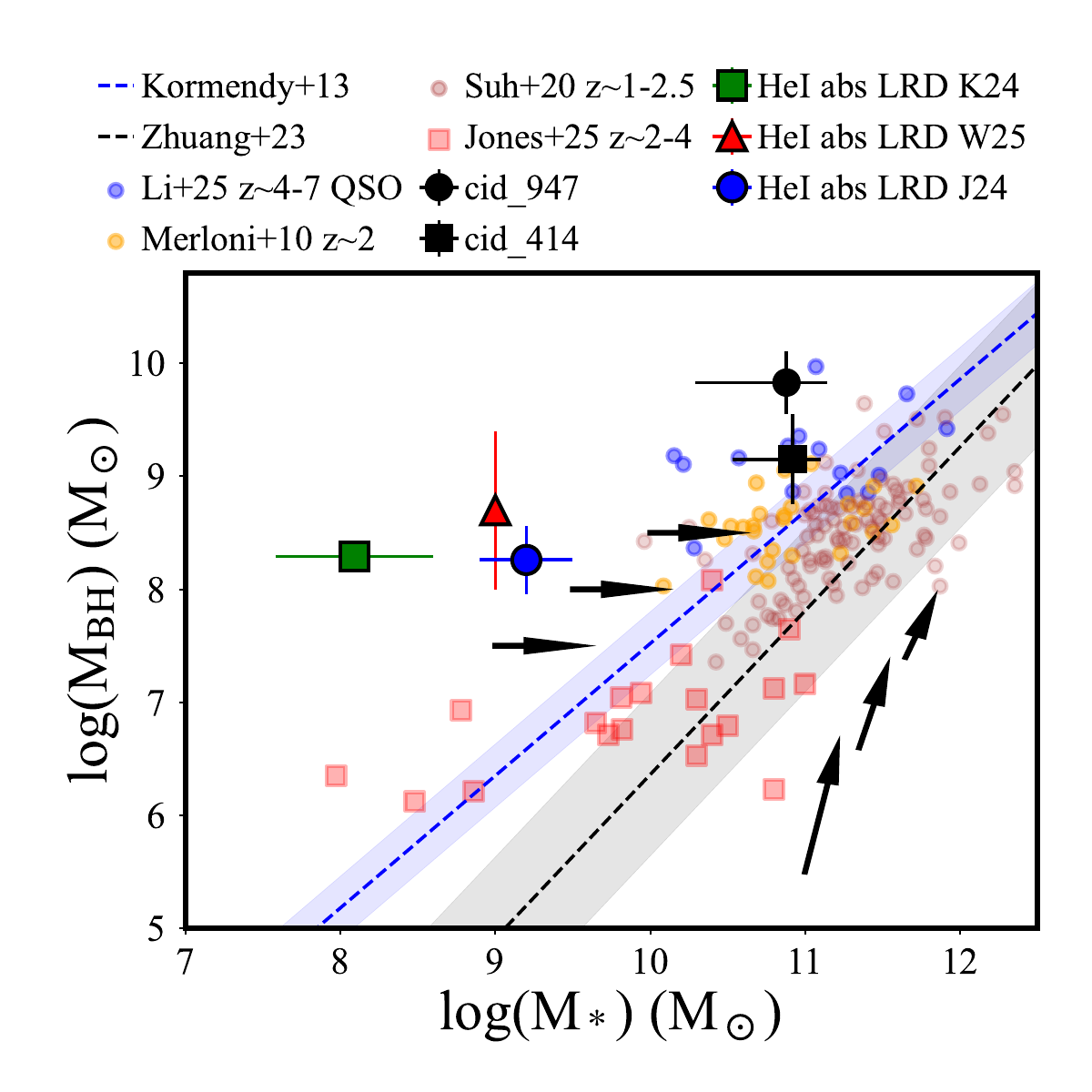}
\caption{\textit{Left}: Host galaxy star formation rate (SFR) and stellar mass ($M_*$) from SED fitting (squares: \cf; circles: \cn). The pink shaded region indicates the star-forming main sequence from \citet{whitaker12}.
\textit{Right:} Stellar mass versus black hole mass. The local relations from \citet{kormendy13} and \citet{zhuang23} are shown as blue and black dashed lines, with their corresponding 1$\sigma$ shaded regions. The black arrows indicate the two black hole–host galaxy growth pathways proposed in \citet{zhuang23}. Blue symbols mark high-luminosity quasars at $z\sim6$ from \citet{galfits}, including sources from ASPIRE (PID \#2078, PI: F. Wang), EIGER (PID \#1243, PI: S. Lilly), and the Subaru High-redshift Exploration of Low-luminosity Quasars survey. The squares represent the JWST AGN compilation \citep{jones25} at $z \sim~2-4$. For comparison, we also include lower redshift AGN samples: optically selected broad line quasars at $z\sim~1-2.2$ \citep{merloni10} and X-ray selected quasars at $z$ up to 2.5 \citep{suh20}}. Three additional AGNs with \hei\ absorption from the literature are also shown: RUBIES 40579 \citep{kocevski25}, W25 \citep{wang25}, and J25 \citep{juodzbalis24}.
\label{fig:sfr_bh_stellar}
\end{figure*}

For both the \texttt{CIGALE} and \texttt{AGNfitter} fitting results, \cf~shows a relatively higher star formation rate (SFR) than \cn, as listed in Table~\ref{table:prop}. We plot the star-forming main sequence in the $M_{*}$–SFR plane in Figure~\ref{fig:sfr_bh_stellar}. Our measurements reveal that \cf~lies on the star-forming main sequence at $z\sim3$, while \cn~is slightly below the relation. 




\subsection{Black hole mass and Eddington ratio measurements}\label{sec:bh_edd}

Since we do not have archival observations of broad Balmer lines or Mg {\sc ii} for \cf, which are typically used to estimate the black hole mass in type I AGNs at similar redshifts in the literature, we instead adopt an indirect empirical scaling relation based on the He {\sc i} and H$\beta$ lines, as calibrated from the local BASS sample \citep{ricci17}. With our measured FWHM of the \hei~emission line and the X-ray luminosity, we obtain a black hole mass of \cf, $\log(M_{\rm BH}/M_{\odot}) = 9.15 \pm 0.40$. The uncertainty arises from the measurement errors in the \hei\ FWHM and the intrinsic scatter of the calibration. For \cn, we adopt the black hole mass $\log(M_{\rm BH}/M_{\odot}) = 9.84 \pm 0.28$ estimated from the broad H$\beta$ line, following the method described in \citet{science}. The bolometric luminosities of the two targets are converted from the 2 to 10 keV luminosity using the bolometric correction from \citet{duras20}. The resulting Eddington ratios are $\lambda_{\rm Edd} = 0.28_{-0.18}^{+0.41}$ for \cf\ and $\lambda_{\rm Edd} = 0.013_{-0.01}^{+0.01}$ for \cn.
Since we do not have archival observations of broad Balmer lines or Mg {\sc ii} for \cf, which are typically used to estimate the black hole mass in type I AGNs at similar redshifts in the literature, we instead adopt an indirect empirical scaling relation based on the He {\sc i} and H$\beta$ lines, as calibrated from the local BASS sample \citep{ricci17}. With our measured FWHM of the \hei~emission line and the X-ray luminosity, we obtain a black hole mass of \cf, $\log(M_{\rm BH}/M_{\odot}) = 9.15 \pm 0.40$. The uncertainty arises from the measurement errors in the \hei\ FWHM and the intrinsic scatter of the calibration. For \cn, we adopt the black hole mass $\log(M_{\rm BH}/M_{\odot}) = 9.84 \pm 0.28$ estimated from the broad H$\beta$ line, following the method described in \citet{science}. The bolometric luminosities of the two targets are converted from the 2 to 10 keV luminosity using the bolometric correction from \citet{duras20}. The resulting Eddington ratios are $\lambda_{\rm Edd} = 0.28_{-0.18}^{+0.41}$ for \cf\ and $\lambda_{\rm Edd} = 0.013_{-0.01}^{+0.01}$ for \cn.

\begin{figure}[ht!]
\epsscale{1.15}
\plotone{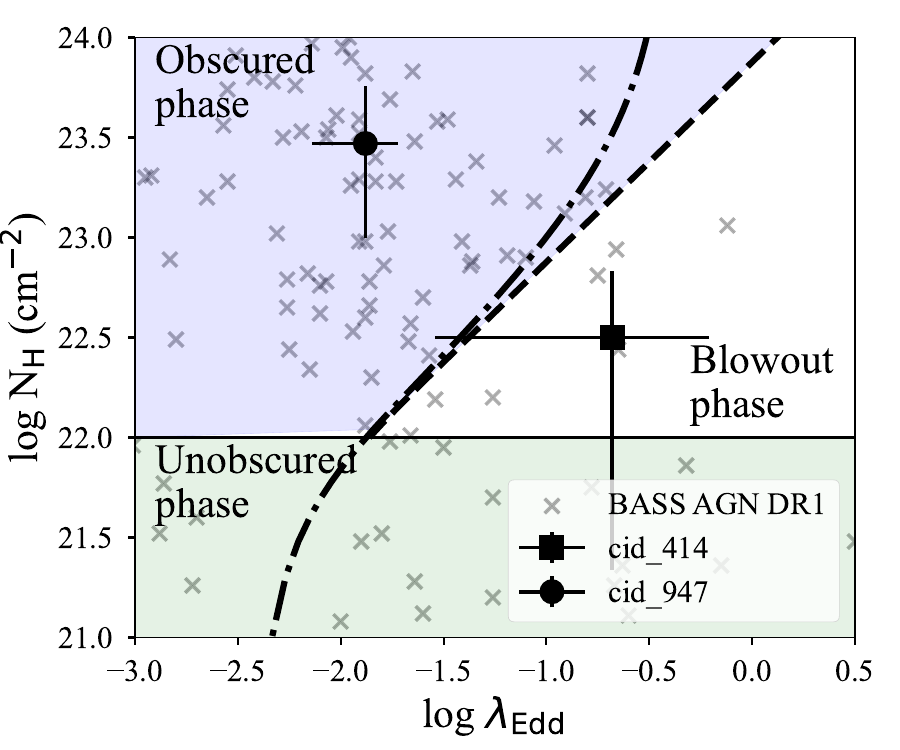}
\caption{X-ray–inferred hydrogen column density versus Eddington ratio. The dashed and dash-dotted curves show the effective Eddington limits for the single-scattering case \citep{fabian09} and for models including IR radiation trapping \citep{ishibashi18}, respectively. The blue shaded region marks the obscured regime, while the green shaded region indicates unobscured AGNs. The white region represents the blowout phase where AGN radiative feedback is expelling the surrounding dusty gas.The grey crosses are AGNs from BASS DR1. Data of column density is from \citet{cricci17} and Eddington ratios are from \citet{koss17}. }\label{fig:nh_edd}
\end{figure}

\section{Discussion}\label{sec:discussion}



\subsection{Dense Gas in Different SMBH Growth Phase}\label{sec:cloudy}

Given the distinct SED shapes, spectral features, and physical properties of the two targets, we tentatively examine the role of dense absorbing gas in AGN obscuration and in the growth of black holes and their host galaxies.

We compare the stellar and black hole masses of the two targets with other AGN populations, including X-ray AGNs at $z = 1
$--$2.5$ \citep{merloni10,suh20}, LRDs with \hei~absorption at $z\sim3$ \citep{wang25,kocevski25,juodzbalis24}, JWST-detected AGNs \citep{jones25}, and quasars at $z\sim4$--$7$ \citep{ruancun25}. Figure~\ref{fig:sfr_bh_stellar} shows the $M_{\rm BH}$–$M_\star$ relation derived from local AGNs \citep{zhuang23,kormendy13}. We find that our two targets, though exhibiting relatively different SED shapes and black hole masses, lie above the empirical relation derived from local AGN samples. Their black hole masses are higher than those of AGNs at $z\sim1$–$2.5$ and are comparable to those of $z\sim5$ quasars, indicative of rapid black hole growth. Between the two targets, \cn~has a relatively higher black hole mass but a lower \edd, which may indicate that \cn~has already undergone a period of black hole growth and that its current activity is suppressed by relatively denser gas (higher $N_{\rm H}$ and $N_{\rm HeI}$). The source \cf, on the other hand, as suggested by \citet{shuqi25}, may represent a later phase of LRD growth and is rapidly accreting material to build up its black hole toward a quasar.

To probe the role of dense gas in obscuration and black hole growth, we place the two targets on the \edd–$N_{\rm H}$ relation from \citet{riccinature17} (Figure~\ref{fig:nh_edd}), derived from 836 local X-ray AGNs \citep{cricci17b,koss17}. In this framework, \edd~regulates nuclear obscuration, with AGNs evolving from an obscured growth phase to a short, unstable blowout phase and finally to an unobscured quasar stage. 

We find that \cf~lies in the ``forbidden region" where radiation pressure is expected to clear the obscuring gas, yet its \hei\ absorption feature does not show a significant velocity offset. Because absorption lines probe only gas along the line of sight, this does not necessarily imply the absence of outflows. Instead, the high-velocity shell may already have been cleared from our viewing direction, leaving behind dense clumps that are more resistant to radiative acceleration or newly infalling material that produces near-systemic absorption. Alternatively, the main outflow may be oriented away from our line of sight, such that the high Eddington ratio drives blowout but no blueshifted absorption is detected in projection.

In contrast, \cn~remains obscured despite its larger black hole mass, suggesting that its main growth phase likely occurred earlier. Strong outflows have also been observed in AGN that remain in the obscured phase \citep{alonso21}, indicating that powerful winds can be launched before the obscuring material is fully cleared. The persistence of obscuration therefore suggests a time delay between the onset of outflows and the removal of the surrounding gas.

We further model the nuclear absorber using \texttt{CLOUDY} C25.00 \citep{c25}, adopting an AGN continuum plus the UV background \citep{mf87}. We vary $\log U$ ($-4$ to $2$), $\log n_{\rm H}$ ($2$ to $10$), and metallicity, fixing $N_{\rm H}$ to the observed value. The best-fit parameters that reproduce the observed \hei~column densities are $\log U = -1.0_{-0.5}^{+0.2}$ and $\log n_{\rm H} = 10.0_{+1.0}^{-0.8}$ for \cf, and $\log U = 0.0_{-1.0}^{+0.3}$ and $\log n_{\rm H} = 9.0_{+2.0}^{-0.8}$ for \cn. The results are insensitive to metallicity between 0.1 and 1 $Z_\odot$. The inferred high gas densities are consistent with predictions for LRDs \citep{kohei25}.

\begin{figure*}[ht!]
\epsscale{1.15}
\plottwo{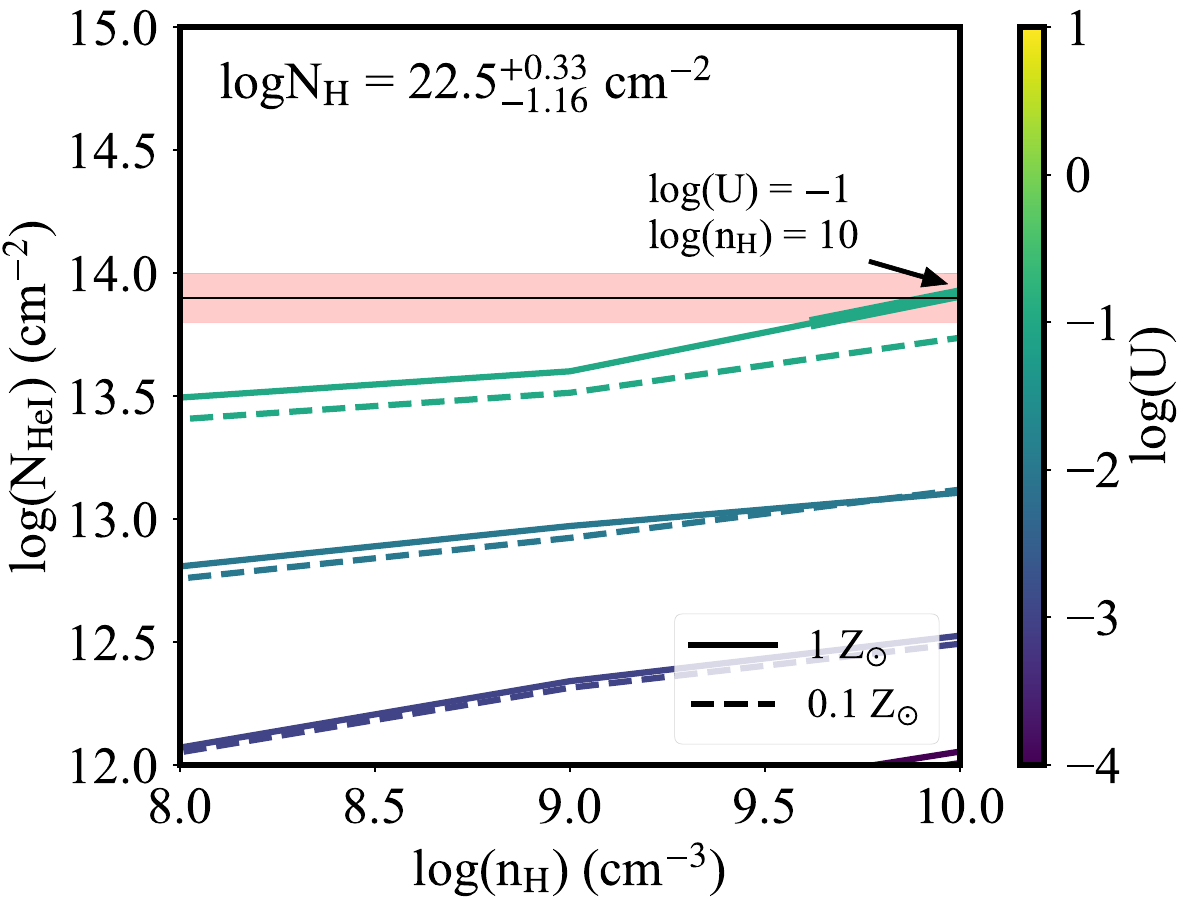}{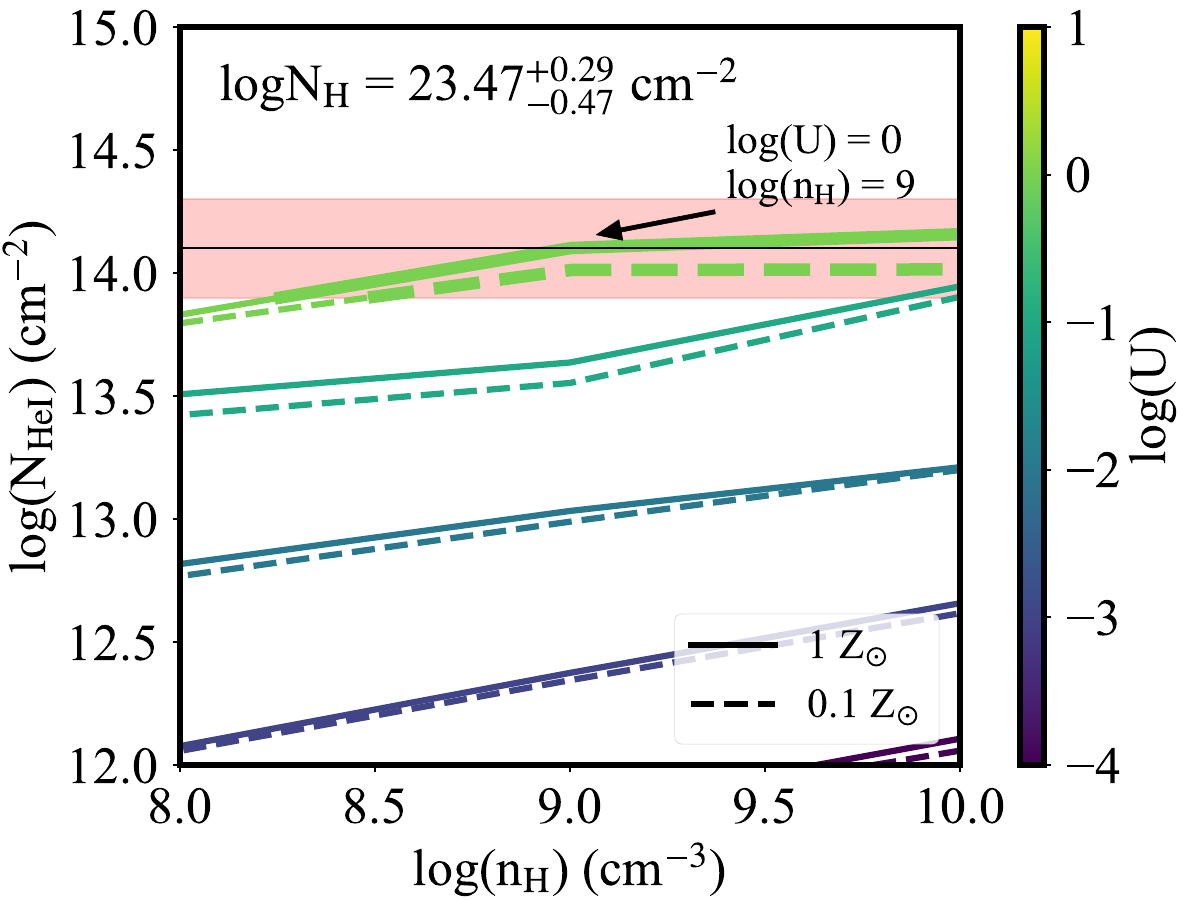}
\caption{The column density of He\,{\sc i} ($\log N_{\rm HeI}$) in the metastable level as a function of gas volume density from the CLOUDY modeling. The left panel shows \cf\ and the right panel shows \cn, each modeled with a different stopping criterion in $N_{\rm H}$. Solid lines represent models with solar metallicity and the dashed lines represent models with $\log(Z/Z_{\odot})=-1$. Colors indicate the value of $\log U$. The horizontal solid line and red shaded region mark the observed $\log N_{\rm HeI}$ and its uncertainty for each source. } \label{fig:cloudy}
\end{figure*}

\subsection{Dust torus and dense absorbing gas geometry}\label{sec:torus}


Given the MIRI detections at $\geq10$–$20~\mu$m and the constraints from the photoionization modelling, we tentatively explore the geometry of the AGN-heated dust and the dense absorbing gas. In the SED fitting (Figure~\ref{fig:sed}), we include dust components modeled as blackbodies with temperatures of 1500~K. For \cf, the MIRI fluxes between 10 and 21~$\mu$m are well reproduced by a $\sim1500$~K component. Together with the prominent outflow signatures, these results suggest that \cf\ is experiencing strong AGN feedback associated with rapid SMBH growth.

The geometry of the dense absorbing gas is likely complex, and our discussion below provides only a simplified estimate. Under a single-slab assumption, the absorber thickness is $l = N_{\rm H}/n_{\rm H}$, giving $l \sim 1.0\times10^{-7}-6.5\times10^{-6}$~pc for \cf\ and $l \sim 9.5\times10^{-7}-6.0\times10^{-4}$~pc for \cn. The distance between the absorber and the ionizing source is estimated using the ionization parameter:
\begin{equation}
r = \sqrt{\frac{L_{\rm bol} f_{\rm ion}}{4\pi n_{\rm H} c U \langle h\nu_{\rm ion} \rangle}},
\end{equation}
where $f_{\rm ion}$ and $\langle h\nu_{\rm ion} \rangle$ are determined by the AGN SED \citep{mf87}. Using our best-fitting values of $U$, we obtain $r \sim 0.007-0.027$~pc for \cf\ and $r \sim 0.002-0.037$~pc for \cn. Combined with the ionization parameter and density inferred from the CLOUDY modelling, the dense gas in \cf\ likely resides very close to the nucleus with a small covering factor. This dense material remains in the line of sight, producing absorption on top of the emission lines even as the system enters a blowout phase. For \cn, the absorbing gas probably resides near the outer BLR, tracing multiple outflowing clumps. These clouds are strongly heated by the AGN radiation field, which lowers the total hydrogen density and produces a higher ionization parameter.

Assuming thermal equilibrium, the inner radius of the dusty torus is
\begin{equation}
R_{\rm torus} = \sqrt{\frac{L_{\rm bol}}{4\pi\sigma T^{4}}},
\end{equation}
where $\sigma$ is the Stefan--Boltzmann constant and we adopt a dust sublimation temperature of $T = 1500$~K \citep[e.g.,][]{zhang15}, yielding $R_{\rm torus} \approx 1.19$~pc for \cf\ and $R_{\rm torus} \approx 0.54$~pc for \cn. Considering the uncertainty in $L_{\rm bol}$ converted from the X-ray $\,2$--$10$~keV luminosity, we also adopt the $L_{\rm bol}$ estimated from the SED fitting described in Section~\ref{sec:sed}. This yields $R_{\rm torus} = 0.86$--$1.19$~pc for \cf\ and $0.54$--$0.96$~pc for \cn. The absorber thickness and distance are smaller than the torus scale and comparable to the BLR size \citep{juodzbalis24}, implying that the absorbing gas lies at BLR-like radii, shares the BLR kinematics, and is influenced by the obscuration geometry of the hot dust torus. 

\subsection{Luminosity function and comparison with other AGN populations }\label{sec:dndz}
We further compare this population of X-ray and MIRI bright AGNs with dense absorbing gas to other populations, including type I quasars, X-ray selected AGNs, and LRDs.

\begin{figure*}[ht!]
\epsscale{0.86}
\plotone{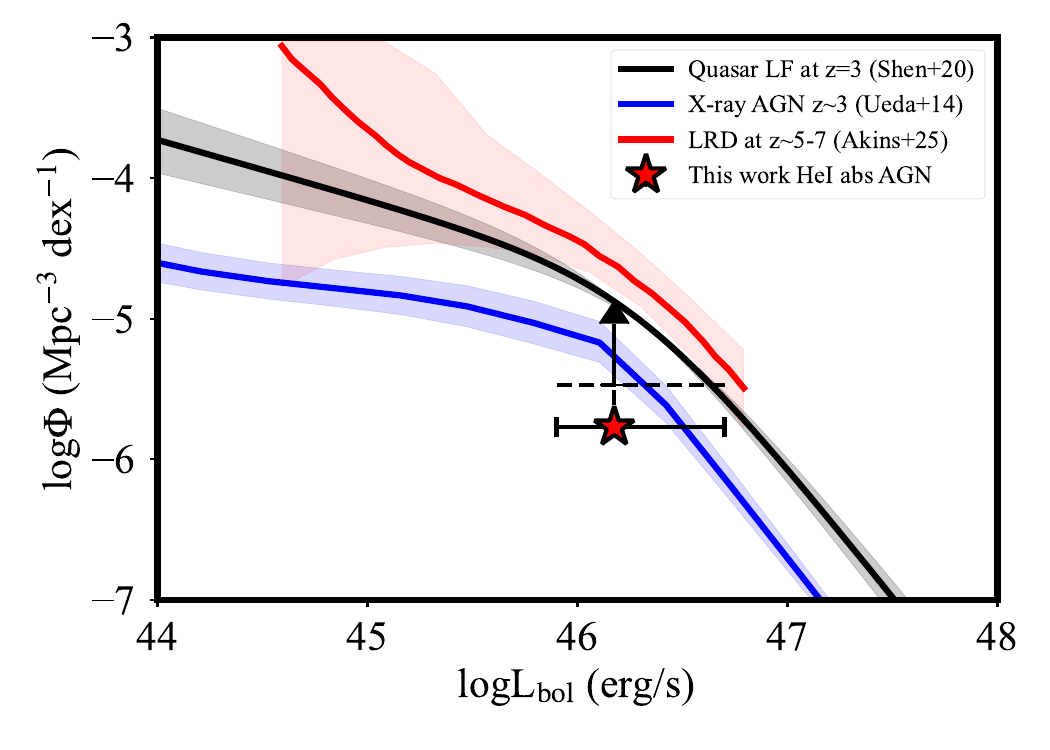}
\caption{The luminosity function (LF) of different AGN populations. The red star is the LF of the \hei\ absorption AGNs reported in this work. The red star and the black dashed line indicates the lower limits, including two additional \hei\ absorption AGNs in the C3D fields with MIRI detections (one presented in \citealt{shuqi25}, and one with very weak He{\sc i} absorption, $\log N_{\rm He\textsc{i}}<12.0$, which is not included in this work). The black line shows the quasar bolometric luminosity function (BLF) at $z=3$ from \citet{shen20}. The green line represents the BLF of the X-ray AGN compilation in \citet{ueda14}. For comparison, we also include the BLF of LRDs at $z\sim5$--$7$ (red line; \citealt{akins}). }\label{fig:lf}
\end{figure*}

With the F444W grism spectral coverage, we are able to detect \hei\ at redshifts $z = 2.7$–$3.6$. 
Besides the two targets analyzed in this work, there are two additional sources in the C3D coverage that exhibit \hei\ absorption and MIRI detections (one reported in \citealt{shuqi25}, and another with very weak \hei~absorption, $\log N_{\rm He\textsc{i}}<12.0$). We do not present them in this work because they did not have secure spectroscopic redshifts prior to the C3D observations, which excludes them based on our selection criteria described in Section \ref{sec:target_selection}. We plot the loosely constrained luminosity function (LF) of the two AGNs at $z \sim 3$ in Figure~\ref{fig:lf}, and the LF including the two additional sources is indicated by the black dashed line. We estimate the luminosity density $\Phi$ following \citet{sun23} as $\Phi = N_{\rm src}/(V_{\rm max},\mathrm{d}\log L)$, where $N_{\rm src}$ is the number of objects in the luminosity bin, $\mathrm{d}\log L$ is the bin width, and $V_{\rm max}$ is the comoving volume (in Mpc$^{-3}$) within the redshift interval $z=2.7$–3.6 and the 0.14 deg$^{2}$ C3D MIRI coverage area. Since no previous \hei~absorption X-ray AGN LF exists at $z\sim3$ and our selection in this work is not strictly complete, we do not apply any completeness correction. Instead, we treat the resulting LF as a lower limit and plot it in Figure~\ref{fig:lf}. The abundance of AGNs with \hei\ absorption and mid-IR detections is about 0.7 dex lower than the predicted abundance of quasars with similar luminosities at $z=3$ (black dashed line; \citealt{shen20}), and about one dex lower than the LRD population at $z=$5–7 \citep{akins}. 

In the pre-JWST era, broad absorption line (BAL) quasars accounted for 10–30\% of the total quasar population \citep{balfrac_radio, balfrac_xray, balfrac}, and most identified BAL systems showed absorption in high-ionization transitions such as C~{\sc iv} and Si~{\sc iv}. JWST has now revealed more AGNs exhibiting \hei\ and/or Balmer-line absorption at high redshift \citep{maiolino23, matthee24}. The fraction of LRDs with Balmer absorption at $z = 4$-$6$ can reach 10–20\% \citep{lin+24}. We note that most current JWST spectra of LRDs with \hei~absorption have been obtained using the low–resolution prism, and the fraction of AGNs exhibiting such absorption may increase as higher–resolution spectroscopy becomes more widely available. 
We note that our estimate is based on only two objects, and a larger sample is required to place stronger constraints on the luminosity function of absorption-line AGNs at this epoch.


\section{Summary}
In this letter, we present two AGNs (\cf\ and \cn) at $z\sim3$ in the COSMOS field that are detected in X-rays and the mid-infrared and exhibit prominent \hei\ absorption. The objects are selected from the COSMOS X-ray AGN catalog \citep{marchesi16b} and are required to be covered by the MIRI observations in the COSMOS-3D survey. The two sources show distinct rest-frame UV–optical SEDs: \cf\ displays an LRD-like V-shaped SED, while \cn\ has a flatter UV–optical slope. A strong \lya\ emission line is detected in \cf\ with no accompanying UV metal emission, whereas \cn\ shows blueshifted \hei\ absorption consistent with outflows traced by C~{\sc iv} and Si~{\sc iv}. These properties suggest that the two AGNs may probe different black hole growth stages or AGN types, while both are embedded in dense gas and luminous in X-rays and the mid-infrared.

We investigate the role of dense absorbing gas in AGN obscuration and black hole growth in these two systems. The source \cf\ has a higher X-ray luminosity, \edd~and lower gas obscuration fraction. 
The $N_{\rm H}$ and \edd~values suggest that \cf\ is in an unstable blowout phase, in which radiation pressure is clearing the obscuring gas, whereas \cn\ remains in a more obscured phase. Photoionization modeling further indicates that the dense absorbing gas is much smaller in scale than the dusty torus and likely resides near the outer regions of the broad-line region. Together with the \hei\ absorbing gas in LRDs detected at $z \sim 3$, we suggest that dense absorbing gas is plausibly common during periods of rapid black hole growth and may play a critical role in regulating AGN obscuration and the co-evolution of black holes and their host galaxies. A larger sample is needed to better constrain the luminosity function and to establish a more complete evolutionary picture of such absorption-line AGNs, including LRDs, at high redshift.



\begin{acknowledgements}
This work is sponsored by the National Key R\&D Program of China (MOST) with grant no.2022YFA1605300. ZJL and SZ acknowledge support from the Chinese Academy of Sciences (no. E5295401). JBC acknowledges funding from the JWST Arizona/Steward Postdoc in Early galaxies and Reionization (JASPER) Scholar contract at the University of Arizona. ZJL, SZ, J-S.H anc C.C. acknowledges support from the National Natural Science Foundation of China (NSFC, grant No.\ 12273051) and the Chinese Academy of Sciences (no. E52H540101). C.C. acknowledges NSFC grant No. 12173045, China Manned Space Program with grant no. CMS-CSST-2025-A07 and Chinese Academy of Sciences South America Center for Astronomy (CASSACA) Key Research Project no. E52H540301. J.-T.S. is supported by the Deutsche Forschungsgemeinschaft (DFG, German Research Foundation) - Project number 518006966.

This work is based on observations made with the NASA/ESA/CSA James Webb
Space Telescope. The data were obtained from the Mikulski Archive for Space Telescopes at the Space Telescope Science Institute, which is operated by the Association of Universities for Research in Astronomy, Inc., under NASA contract NAS 5-03127 for JWST. These observations are associated with programs \#5893. Support for program \#5893 was provided by NASA through a grant from the Space Telescope Science Institute, which is operated by the Association of Universities for Research in Astronomy, Inc., under NASA contract NAS 5-03127. Some/all of the data presented in this article were obtained from the Mikulski Archive for Space Telescopes (MAST) at the Space Telescope Science Institute. The specific observations analyzed can be accessed via \dataset[doi: 10.17909/ys3r-yp43]{https://doi.org/10.17909/ys3r-yp43}. We acknowledge the strong support provided by the program coordinator Alison Vick and instrument reviewers Brian Brooks and Jonathan Aguilar. Support for this work was provided by NASA through grant JWST-GO-01727 awarded by the Space Telescope Science Institute, which is operated by the Association of Universities for Research in Astronomy, Inc., under NASA contract NAS 5-26555. This research employs a list of Chandra datasets, obtained by the Chandra X-ray Observatory, contained in~\dataset[DOI: 10.25574/cdc.543]{https://doi.org/10.25574/cdc.543}

\end{acknowledgements}

\facilities{JWST, HST, Chandra, Keck:I (LRIS), Keck:II (DEIMOS), ALMA, Herschel, VLA}

\bibliography{sample701}{}
\bibliographystyle{aasjournalv7}

\appendix
\restartappendixnumbering

\section{Figures}



\begin{figure*}[ht!]
\includegraphics[width=0.48\linewidth]{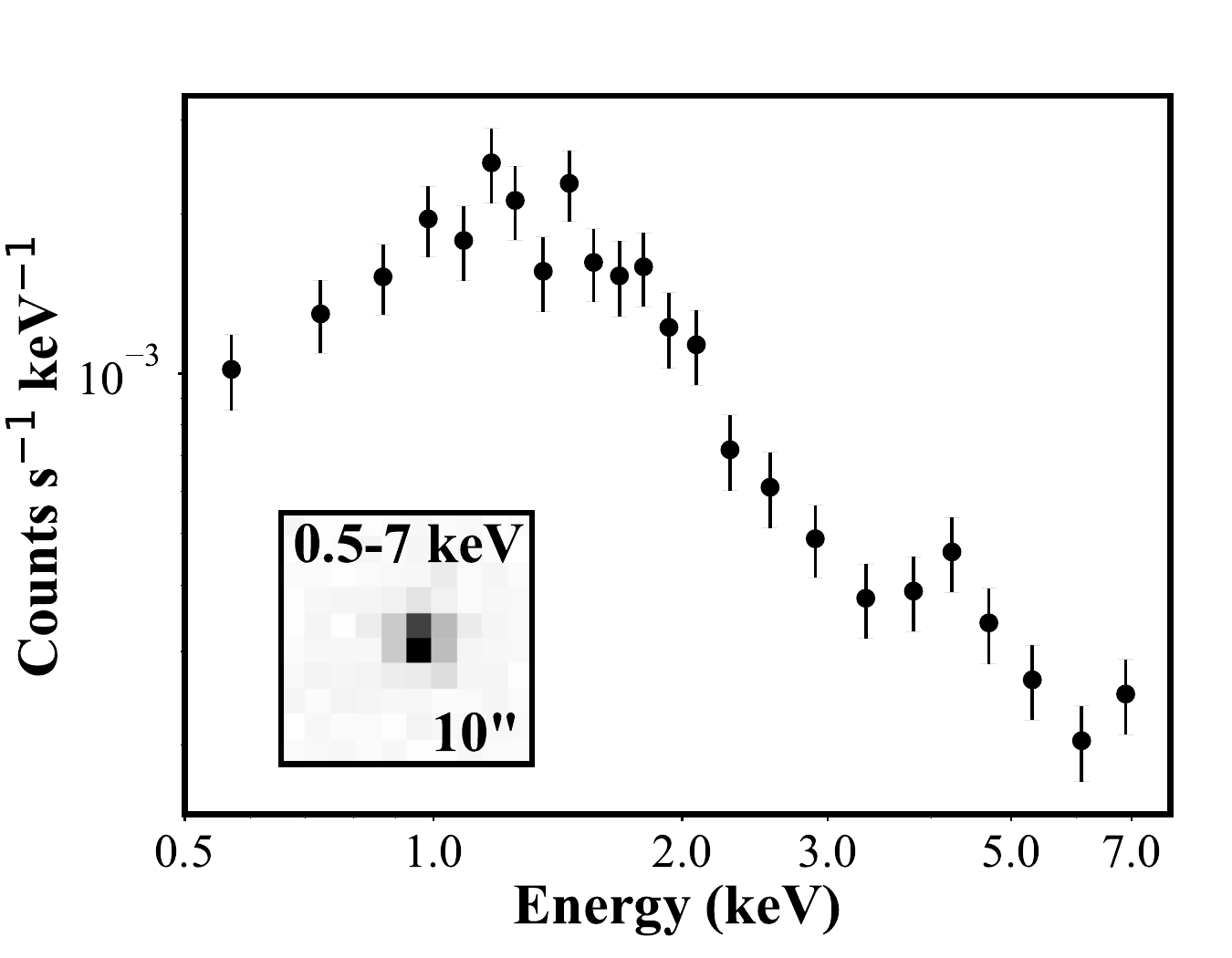}
\includegraphics[width=0.48\linewidth]{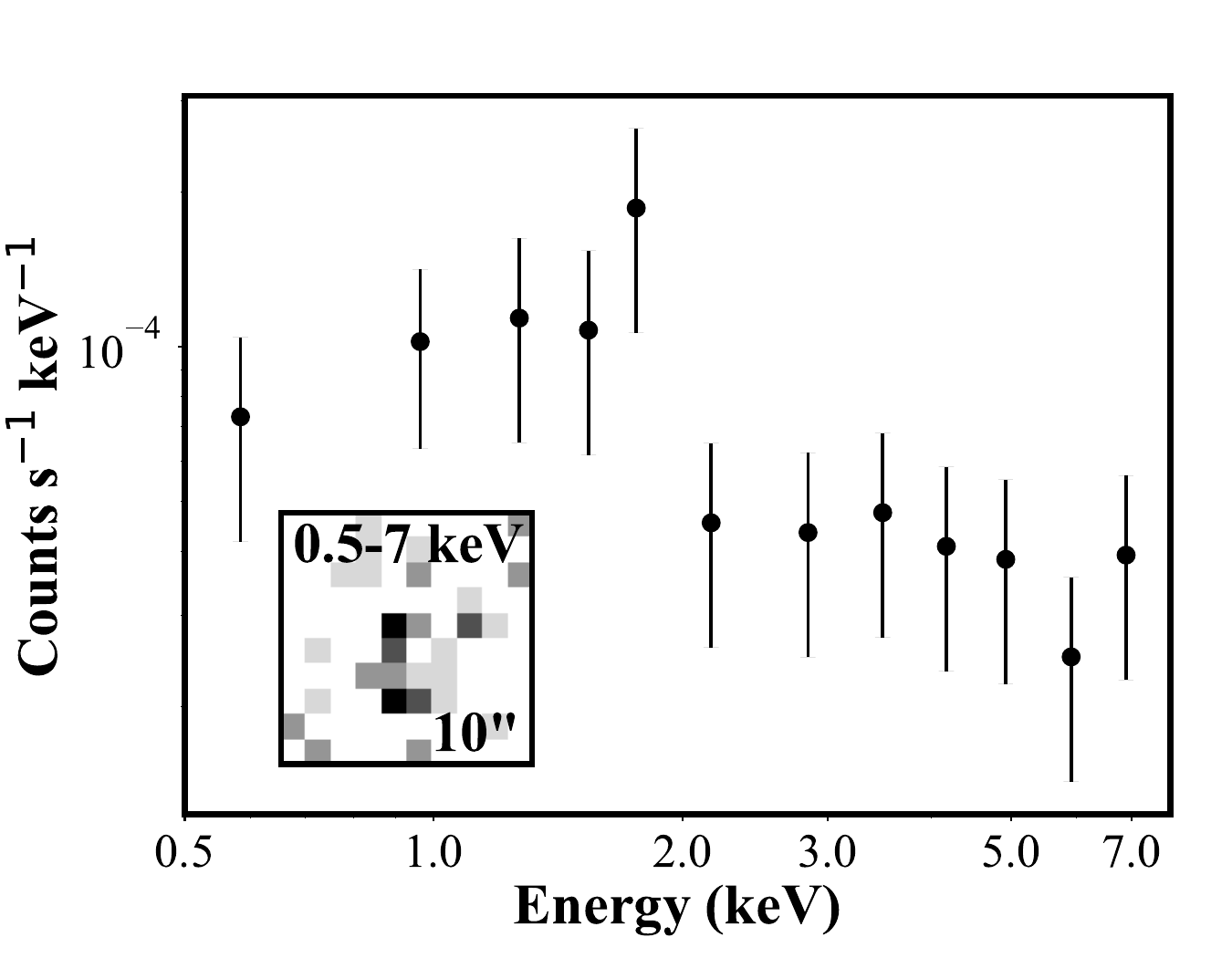}
\caption{0.5-7 keV X-ray imaging and spectroscopy for \cf\ (\textit{left}) and \cn\ (\textit{right}). The X-ray observations are from Chandra COSMOS Legacy survey \citep{civano16}. We use \texttt{CIAO} 4.17 \citep{ciao} to reduce the Chandra data with a 4.12.2 version of calibration database (\texttt{CALDB}). 
The data are reprocessed by the script \texttt{chandra\_repro}. 
The background flares are removed by the command \texttt{deflare}. 
The X-ray imaging is generated from a merged event list by \texttt{merge\_obs}. 
We extract the spectrum in the region enclosing 90\% PSF at 1 keV for each observation with \texttt{specextract} script.
The spectra are then combined and grouped to have at least 10 counts in each energy bin.}
\label{fig:xray_data}
\end{figure*}

\begin{figure*}[ht!]
\epsscale{1.2}
\plotone{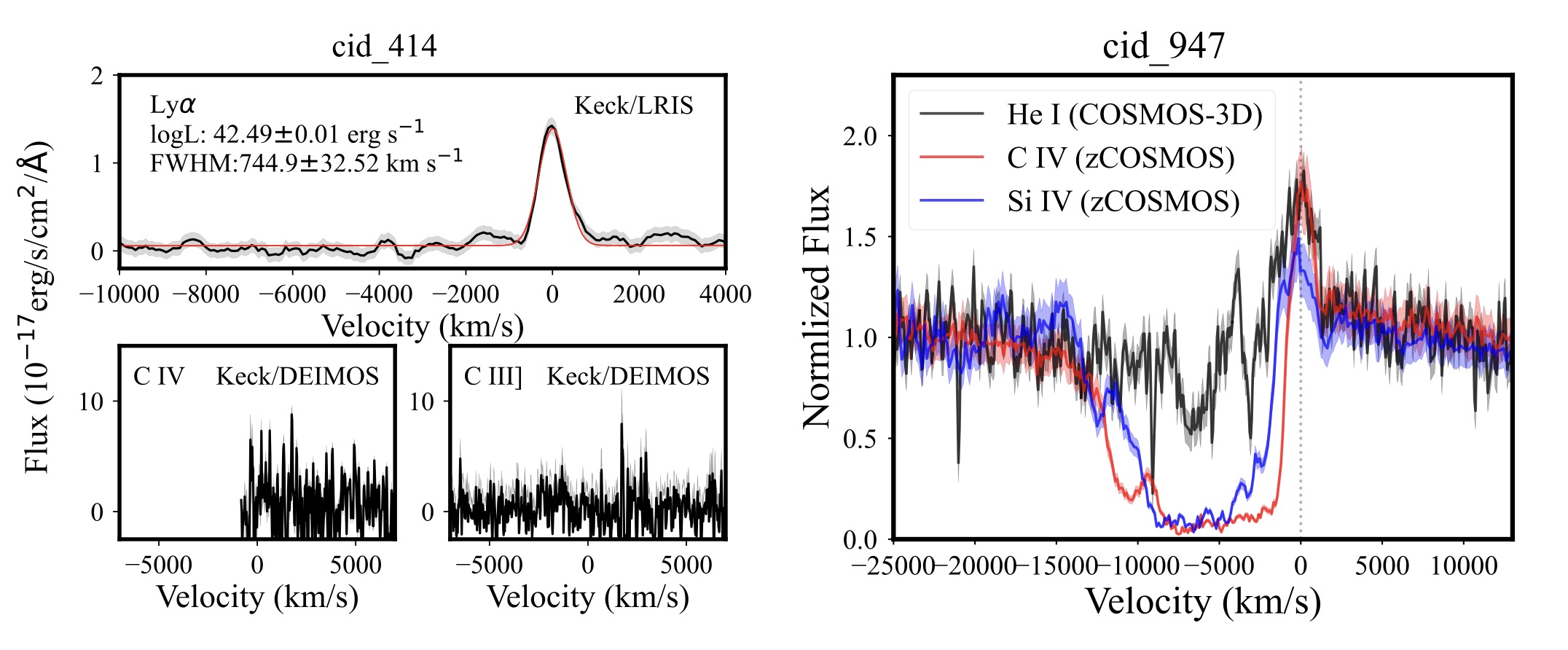}
\caption{Archival Keck spectra for \texttt{cid\_414} (left) and \texttt{cid\_947} (right). A clear \lya~emission line is detected for \texttt{cid\_414} in the CLAMATO survey \citep{clamato,clamato2}, while no C~{\sc iv} or C~{\sc iii}] emission is seen in the Keck/DEIMOS spectrum with an exposure time of 1800 s. In the right panel, we show the velocity profiles of the broad-absorption AGN \texttt{cid\_947}, including \hei, C~{\sc iv}, and Si~{\sc iv}.}
\label{fig:optical_spec}
\end{figure*}

\begin{figure}
\includegraphics[width=0.57\linewidth]{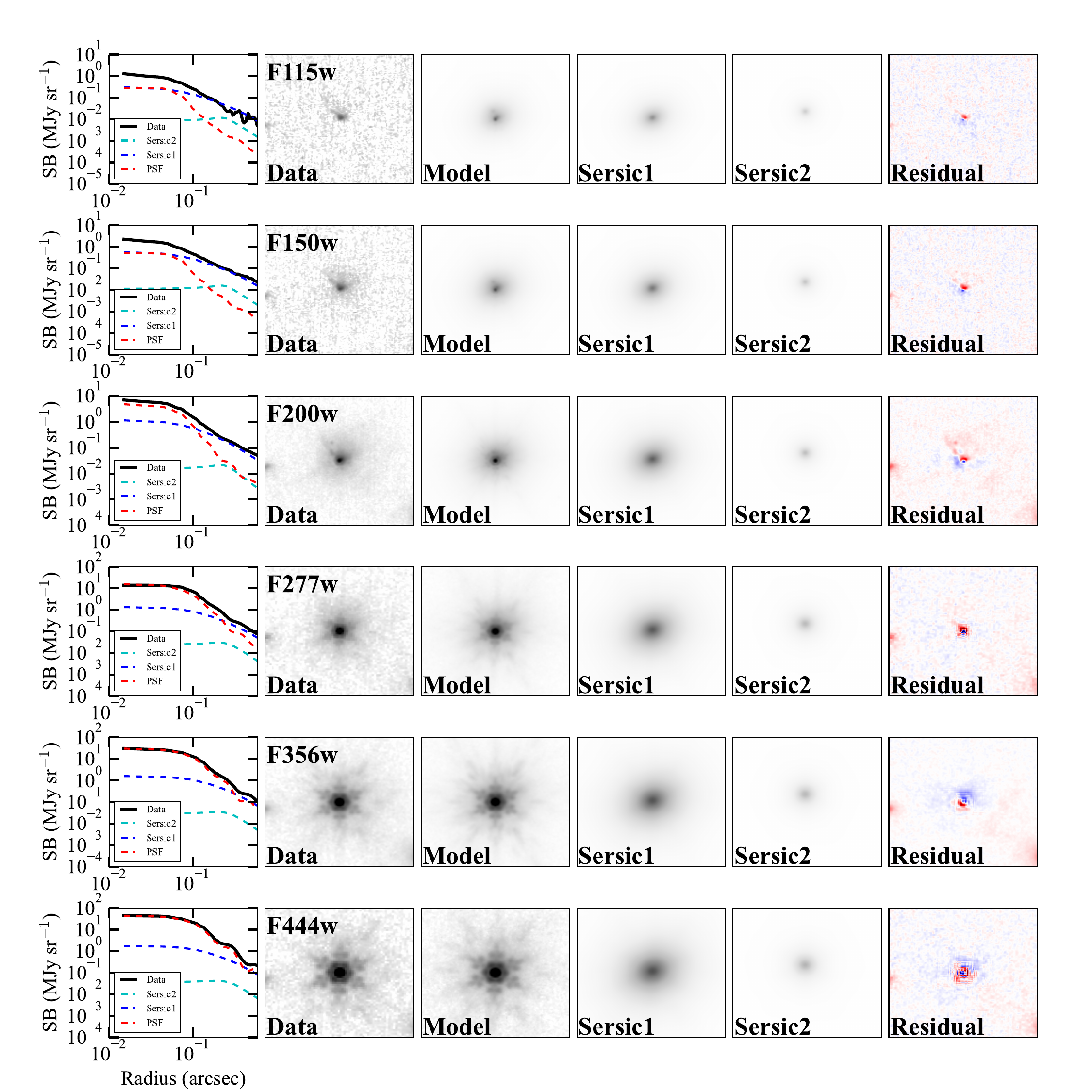}\hspace{-1em}
\includegraphics[width=0.475\linewidth]{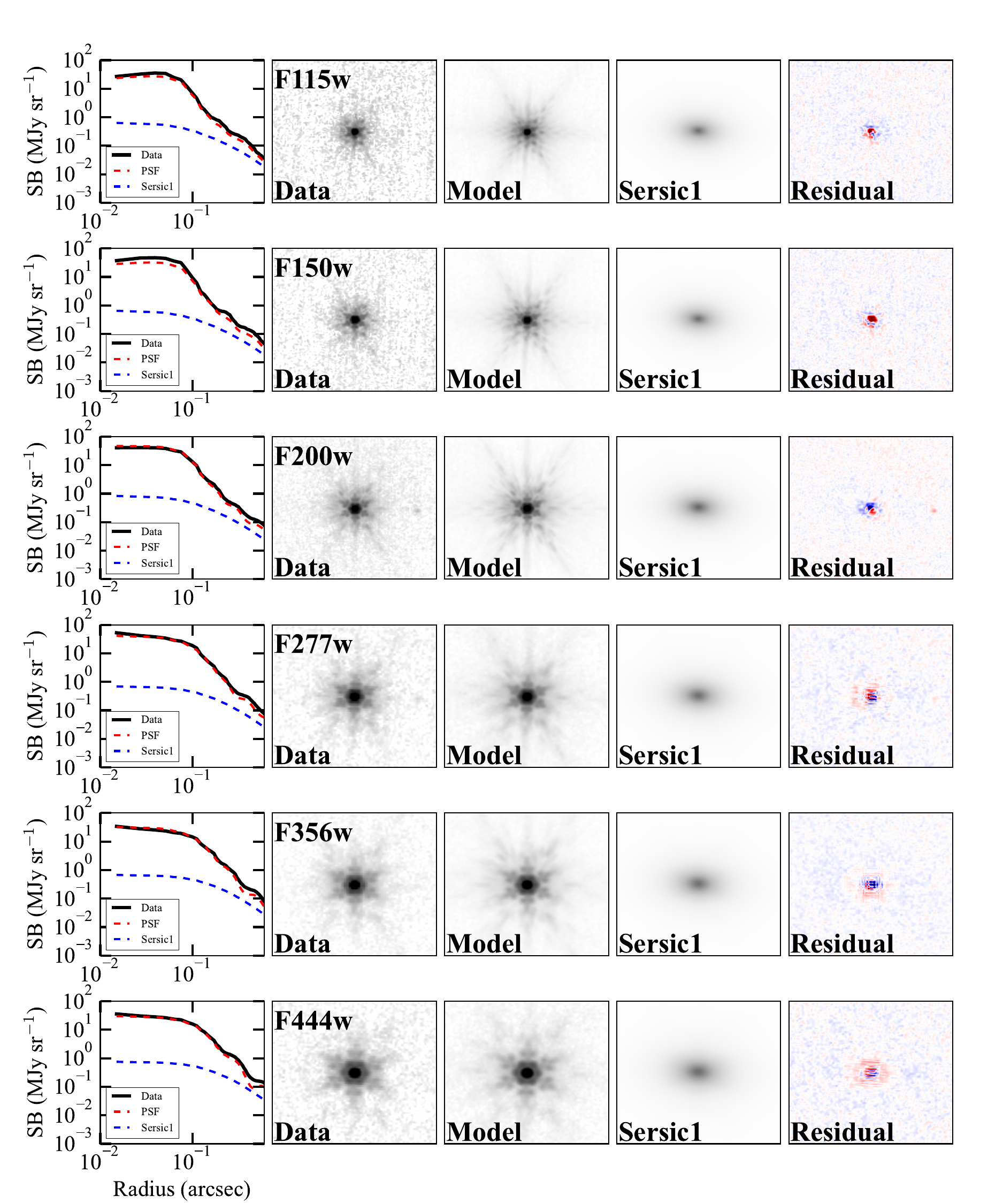}
\caption{Multiband JWST imaging decomposition for \cf\ (left) and \cn\ (right) using \texttt{GALFITS} \citep{galfits}. The 1D surface-brightness (SB) profiles as a function of radius are shown in the first panel of each row. The data and model images are displayed using the same logarithmic stretch, while the residual (data--model divided by the sigma image) is shown linearly over a range of $-10$ to $10$. We model \cf\ with two S\'ersic components. In the F115W and F150W bands, the SB of the second S\'ersic component exceeds that of the first beyond $\sim0\farcs1$.}
\label{fig:galfits}
\end{figure}



\end{document}